%% file: rchargedbh.tex
\documentstyle[prl,aps,preprint,tighten,floats,aps,epsf,psfig]{revtex}
\input lmacros
\begin{document}
\draft
\date{\today}
\preprint{\vbox{\baselineskip=12pt
\rightline{HUTP-{\sl A}009}
\vskip0.2truecm
\rightline{UPR-0832-T}
\vskip0.2truecm
\rightline{hep-th/9902195}}}

\title{Phases of  R-charged Black Holes,  
 Spinning Branes  and Strongly
Coupled Gauge Theories} 
\author{Mirjam 
Cveti\v{c}$^{(1)}$,  and Steven S.~Gubser$^{(2)}$} 
\address{%\small
(1)  David  Rittenhouse  Laboratories, 
University  of  Pennsylvania,
Philadelphia,  PA  19104 
\\
%\small
(2) Lyman Laboratory of Physics, Harvard University,
Cambridge, MA 02138}

\maketitle
\begin{abstract}
 We study the thermodynamic stability of charged black holes in gauged
supergravity theories in $D=5$, $D=4$ and $D=7$.  We find explicitly the
location of the Hawking-Page phase transition between charged black holes
and the pure anti-de Sitter space-time, both in the grand-canonical
ensemble, where electric potentials are held fixed, and in the canonical
ensemble, where total charges are held fixed.  We also find the explicit
local thermodynamic stability constraints for black holes with one non-zero
charge. In the grand-canonical ensemble, there is in general a region of
phase space where neither the anti-de Sitter space-time is dynamically
preferred, nor are the charged black holes thermodynamically stable.  But
in the canonical ensemble, anti-de Sitter space-time is always dynamically
preferred in the domain where black holes are unstable.

We demonstrate the equivalence of large R-charged black holes in $D=5$,
$D=4$ and $D=7$ with spinning near-extreme D3-, M2- and M5-branes,
respectively.  The mass, the charges and the entropy of such black holes
can be mapped into the energy above extremality, the angular momenta and
the entropy of the corresponding branes.  We also note a peculiar
numerological sense in which the grand-canonical stability constraints for
large charge black holes in $D=4$ and $D=7$ are dual, and in which the
$D=5$ constraints are self-dual.

\end{abstract}
%\vskip 0.3truecm
%\pacs{\tt PACS number(s): }
\newpage
\section{Introduction}
\label{Intro}

Since the work of \cite{gkPeet} on the thermodynamics of near-extreme
D3-branes, there has been considerable interest in the extent to which
supergravity solutions reflect the thermal properties of non-abelian gauge
theory in $3+1$ dimensions with ${\cal N}=4$ supersymmetry.  The conformal
invariance of the underlying theory suggests that the only part of the
supergravity solution that is relevant is the near-horizon region, which
amounts to a black hole in five-dimensional anti-de Sitter space ($AdS_5$)
times a five-sphere ($S^5$).  Indeed, following the conjecture
\cite{JuanAdS} that string theory on $AdS_5 \times S^5$ is physically
identical to ${\cal N}=4$ gauge theory on the boundary of $AdS_5$, it was
shown \cite{witHolTwo,WitSuss} that the thermodynamics of large
Schwarzschild black holes in $AdS_5$ matches the expected thermodynamics of
the gauge theory, at least up to constants.  In part this amounts to a
rephrasing of \cite{gkPeet}; but also \cite{witHolTwo} went on to propose a
description of pure $\hbox{QCD}_3$ in terms of the finite-temperature
theory on a near-extreme D3-brane.  A key ingredient to the
confinement-deconfinement transition described in \cite{witHolTwo} was that
the gauge theory is on a three-sphere of finite volume.  The finite volume
breaks the conformal invariance which would otherwise make a phase
transition impossible.

In \cite{Gubser}, the study of near-extreme D3-brane thermodynamics was
extended to include the case where the branes have angular momentum in a
plane orthogonal to the branes.  We call this type of angular momentum
spin.  Like most charged black hole solutions, the D3-brane with no spin
exhibits positive specific heat sufficiently close to extremality.  This
property has long been regarded as the {\it sine qua non} for comparison
with field theory, since field theories are always expected to have
positive susceptibilities.  The surprise in \cite{Gubser} was that even in
the near-horizon limit where field theory is supposed to be applicable, the
thermodynamics is not always stable.  If the spin is too large compared to
the energy, the near-extreme brane solution no longer represents even a
local maximum of the entropy; the entropy fails to remain a subadditive
function of external variables such a energy and angular momentum.  The
stability analysis has been generalized to the M2- and M5-branes, and to
multiple angular momenta \cite{CvGu}.  For D3-, M2- and M5-branes, Cai and
Soh \cite{Cai} have discussed the critical behavior of specific heats and
other susceptibilities near the boundary of stability, and also at other
points in phase space related to stability with fixed angular momenta.

Subadditivity of the entropy is a generalization of the requirement of
positive specific heat.  It amounts to the ``effective'' Euclidean action
for the black hole configuration being a local minimum.  Thus the stability
criterion used in \cite{Gubser} is rather different from \cite{witHolTwo},
where the location of the confinement-deconfinement transition was found by
comparing the Euclidean action of two topologically distinct geometries
which made competing contributions to the path integral.  One geometry
represented a black hole, while the other represented a gas of particles in
anti-de Sitter space.  The discontinuous transition between the two was
first studied by Hawking and Page \cite{HP}.

One of the goals of the present paper is to compare the Hawking-Page
transition to the subadditivity condition on the entropy, for charged black
holes in anti-de Sitter space, i.e., charged black hole configurations in
maximally supersymmetric gauged supergravity~\cite{romans,BCS,Duff}.  In a
realization of the gauged supergravity theories as a truncation of string
theory or M~theory compactified on a sphere, the gauge group is the
isometry group of the sphere.  In the AdS/CFT correspondence this gauge
group becomes the R-symmetry group of the boundary CFT which is thought to
be dual to string theory on the anti-de Sitter geometry.  (Recall, the
R-symmetry algebra is by definition that part of the supersymmetry algebra
which commutes with the Poincare symmetries but not the supersymmetries.
It is like a bosonic flavor symmetry, but evades the Coleman-Mandula
theorem by nevertheless participating in the same superalgebra as the
Lorentz group generators).  We therefore refer to this type of charge as
R-charge.  Our work extends that of~\cite{Gubser} in several ways.  First,
we show that the Kaluza-Klein reduction of near-extreme spinning branes in
fact corresponds to large R-charged black holes in $D=5$, $4$ and $7$.
Their mass and charges correspond to the energy above extremality and the
angular-momenta of the near-extreme spinning D3-, M2- and M5-branes. For
general R-charged black holes we find explicitly the constraints for the
Hawking-Page phase transition as well as subadditivity conditions in the
case of a single R-charge turned on.  We do the analysis with the
temperature and the R-charges fixed, and also with the temperature and the
electric potentials fixed.  The electric potentials are the thermodynamic
dual variables to the R-charges.  We refer to the first type of analysis as
stability in the canonical ensemble, and to the second type as stability in
the grand-canonical ensemble.  A subtlety regarding the differences between
ensembles and the relevance of thermodynamic variables versus fixed
parameters is explained in section~\ref{LocalGlobal}.

The general conclusion from local stability analyses is that branes which
spin too fast, or equivalently black holes with excessive R-charges, are
unstable.  This begs the question, what do such configurations break into?
If the conserved R-charges are regarded as fixed parameters (canonical
ensemble), then in view of our results it seems plausible that excessively
R-charged black holes in anti-de Sitter space undergo a Hawking-Page
transition to a gas of particles.  Our analysis shows that this is indeed
the case.  The transition is an essentially quantum phenomenon which
eliminates an event horizon and a singularity from the space-time.  On the
other hand, if instead of the total R-charges their dual potentials are
specified at the boundary of anti-de Sitter space (grand-canonical
ensemble), we find that the Hawking-Page transition cannot altogether
explain the instability of excessively charged solutions.  We show that for
the grand-canonical ensemble there is always a domain of large R-charges
for which the black hole is dynamically preferred in the sense of Hawking
and Page, but is locally thermodynamically unstable.  That is, the black
hole's Euclidean action is smaller than that of the gas of particles in
anti-de Sitter space, but nevertheless is only a saddle point and not a
local minimum of the action.

In Section~\ref{solution} we exhibit the R-charged black hole solutions in
$AdS_5$, found in \cite{BCS}, which will be the focus of the paper.  In
Section~\ref{Phases} we describe the conditions for local thermodynamic
stability and for the Hawking-Page-type transition.  In
Section~\ref{RCharge} we demonstrate explicitly that spinning near-extreme
D3-branes can be Kaluza-Klein reduced to large R-charged black holes in
$AdS_5$, and show (Subsection~\ref{equiv}) how the thermodynamics of
spinning D3-branes can be recovered from our results in
Section~\ref{Phases}.  In Sections~\ref{FourDim} and~\ref{SevenDim} we
extend our results to charged black holes in $AdS_4$~\cite{Duff} and
$AdS_7$, respectively.  In Section~\ref{Mbranes} we relate these black
holes to spinning M2-branes and M5-branes.

\section{Black Holes in Five Dimensional Gauged Supergravity}
\label{solution}

In this section we summarize the black hole solution in $D=5$ $N=8$ gauged
supergravity. This solution was found in~\cite{BCS} as a special case
(STU-model) of the solutions of $D=5$ $N=2$ gauged supergravity equations
of motion.  The relevant bosonic part of the Lagrangian can be cast in the
following form~\cite{BCS}:

\begin{eqnarray}
e^{-1} \mathcal{L} &=& {\frac{1}{2\kappa^2}} R + g^2{\cal V} - 
{\frac{1}{4}} G_{ij} F_{\mu\nu}
{}^i F^{\mu\nu j}-{\frac{1}{2}}
 { G}_{ij} \partial_{\mu} 
X^i \partial^\mu
X^j +{\frac{e^{-1}}{48}} \epsilon^{\mu\nu\rho\sigma\lambda}\epsilon_{ijk}
F_{\mu\nu}^iF_{\rho\sigma}^jA_\lambda^k \ .   \nonumber \\
\label{action}
\end{eqnarray}
 The space-time indices $\mu$, $\nu$, etc. run from $0$ to $4$.  The metric
has signature -++++.  $R$ is the scalar curvature, $F_{\mu\nu}^i$
($i=1,2,3$) are the Abelian field-strength tensors, $e=\sqrt{-g}$ is the
determinant of the F\"unfbein $e_m{}^a$, and ${\cal V}$ is the scalar
potential given by \cite{BCS}: 
  \begin{equation} 
   {\cal V}=2\sum_{i=1}^3 {1\over X^i} 
  \end{equation} 
 The metric on the relevant part of the scalar manifold is $G_{ij}={1\over
2}{\rm diag}((X^1)^{-2},(X^2)^{-2},(X^3)^{-2})$.  Here $X^i$ are three real
scalar fields $X^i$, subject to the constraint $\prod_{i=1}^3 X^i=1$.  The
solution is specified by the following metric: 
  \begin{equation} 
   ds^2 = -{(H_1 H_2 H_3)^{-2/3}} f dt^2 + (H_1H_2H_3)^{1/3}
 	\bigg(f^{-1} {dr^2} + 
	r^2 d\Omega_{3,k} \bigg) \ , \label{metric}	
  \end{equation}
 where	
 \begin{equation}
f = k - {\mu \over r^2} + g^2 r^2 H_1H_2H_3 \ , \ \ H_i=1+{{q_i}\over r^2}, \
\ i=(1,2,3)\ ,  
\label{f}\end{equation}
 and $d\Omega_{3,k}^2$ is the metric on $S^3$ with unit radius if $k=1$, or
the metric on ${\bf R}^3$ if $k=0$.  $g\equiv {1\over L}$ is the inverse
radius of $AdS_5$, and related to the cosmological constant $\Lambda=
-6g^2=-{6\over L^2}$. The three real scalar fields $X^i$
and  three  gauge field potentials  $A_\mu^i$ are of the form
\begin{equation}
X^i = H_i^{-1}{(H_1 H_2 H_3)^{1/3} }\ , \ A_t^i= {{\tilde q}_i\over{r^2+q_i}} \
, i=1,2,3 \ .
\end{equation}
 Above, we have chosen the Newton's constant $G_5={\pi\over
4}$.\footnote{This choice allows for the most efficient parameterization of
the the physical charges $Q_i={{\Omega_{D-2}(D-3)}\over {16\pi G_D}}{\tilde
q}_i={\tilde q}_i$ ($D=5$).
Our normalization of the gauge fields
differs from that introduced in \cite{BCS}  by a factor of ${2\over L}$.}
Here  the following parameterization for $q_i$ (the ``charges'' entering the 
metric) and ${\tilde q}_i$ (the physical charges) is introduced:
\begin{equation}
q_i = \mu \sinh^2\beta_i \qquad , \qquad 
\tilde q_i =  \mu \sinh \beta_i \cosh\beta_i \ .
\label{charges}\end{equation}

In the following we shall concentrate on the  black hole solutions  with $k=1$.
 The  solution is  specified by the ADM mass:
\begin{equation}
M=\textstyle{3\over 2} \mu +\sum_{i=1}^3q_i=
\textstyle{3\over 2}\left(r_+^2+r_+^4g^2\prod_{i=1}^3H_i\right)+\sum_{i=1}^3q_i \ , 
\label{mass}\end{equation}
where  the second equality follows from the definition  of  $r_+$, the
 radial
coordinate  at the outer horizon,  
  as the largest
non-negative zero of the function $f(r)$ appearing in \eno{f}.

When all the three charges are turned it is not possible to obtain the
explicit form of the entropy $S$ and the inverse temperature $\beta$ in
terms of physical charges ${\tilde q}_i$ and the ADM mass $M$.  However, it
turns out that the entropy $S$ and the inverse Hawking temperature $\beta$
can be expressed explicitly in terms of the ``charges'' $q_i$ and
the radial coordinate of the outer horizon $r_+$:
 \begin{equation}
S= {A\over {4G_N}}=2\pi \sqrt{\prod_{i=1}^3 (r_+^2+q_i)}\ , \label{SDef}
\end{equation}
\begin{equation}
\beta={1\over T_H}={{2\pi r_+^2\sqrt{\prod_{i=1}^3
(r_+^2+q_i)}}\over{2r_+^6+r_+^4(1+\sum_{i=1}^3q_i)-\prod_{i=1}^3q_i}} \ .
\label{beta}\end{equation}
 Again, note that $\tilde{q}_i$ rather than $q_i$ are the physical charges
(i.e., the conserved charges to which Gauss's law applies).  One can express
\begin{equation}
{\tilde q}_i^2=q_i(r_+^2+q_i)\left[1 +{1\over r_+^2}\prod_{j\ne
i}(r_+^2+q_j)\right] \ .
\label{pcharge}\end{equation}
 Also, the electric potentials at the outer horizon take the following form:
\begin{equation}
\phi_i\equiv A^i_t(r_+)={{\tilde q_i}\over {r_+^2+q_i}} \ . 
\label{phi}\end{equation}
 The parameterization \eno{SDef}, \eno{beta}, \eno{pcharge}, \eno{phi} will
turn out to be useful when deriving the the thermodynamics stability
constraints and constraints for the Hawking-Page phase transition for these
configurations.

In the case of a single  non-zero charge, say, $q_1=q\ne 0$, the   horizon
coordinate $r_+$ and $q$  can be expressed explicitly in
terms of   the ADM  mass $M$ and  physical charge ${\tilde q}$, and thus
 all the corresponding thermodynamic quantities $S$, $\beta$ and $\phi$ can
be written explicitly in terms of $M$ and ${\tilde q}$.  The following
relationships are useful in this case:
  \begin{eqnarray}
   r_+^2 &=& {-1-q + \sqrt{(1+q)^2 + 4\mu} \over 2} \ ,  \nonumber\\
   q &=& \sqrt{\tilde{q}^2 + {\mu^2 \over 4}} - {\mu \over 2} \ , \nonumber \\
   \mu &= &{4 \over 3} M - {2 \over 3}
    \sqrt{M^2 + 3 \tilde{q}^2} \ .
  \end{eqnarray}
 Using these expressions, one can express the entropy $S=2\pi
r_+^2\sqrt{r_+^2+q}$ wholly in terms of $M$ and $\tilde{q}$.

Before we proceed, a comment on the choice of units is in order.  It is not
completely general to choose units where simultaneously $G_5 = {\pi\over
4}$ and $g\equiv {1\over L}= 1$. ($L$ is related to the asymptotic cosmological
constant $\Lambda=-{6\over L^2}$.) The reason is that $G_5$ is a length
cubed and $g$ is an inverse length, so $G_5 g^3$ is a pure number.  If $N$
is the number of units of Dirac flux supporting the $AdS_5 \times S^5$
geometry, then one finds $G_5 g^3 = {\pi \over 2 N^2}$.  So picking $G_5 =
{\pi\over 4}$ and $g = 1$ simultaneously actually fixes $N = \sqrt{2}$.

Nevertheless, in the following sections we shall take $G_5={\pi\over 4}$
{\it and} $L\equiv {1\over g}=1$.  Powers of $L$ can be restored by making
the following replacements: $r_+ \to {r_+\over L}$, $q_i \to {q_i\over
L^2}$, $\mu \to {\mu\over L^2}$, $M \to {M\over L^2}$, and $\tilde{q}_i \to
{\tilde{q}_i\over L^2}$.  In Section~\ref{RCharge}, where we relate the
spinning D3-branes and their thermodynamics to that of large black holes,
we restore $L$ explicitly.

\section{Phases of R-charged Black Holes}
\label{Phases}

\subsection{Ensembles and the nature of stability}
\label{LocalGlobal}

We will explore two notions of thermodynamic stability: local and global.
Roughly speaking, the distinction is local versus global minima of an
appropriate thermodynamic potential.  There are various subtleties to
explain, and we will address them in this section.

Local thermodynamic stability is the statement that the entropy $S$ as a
function of the other extensive thermodynamic variables $x_i$ is
subadditive in a sufficiently small neighborhood of a given point in the
phase space of possible $x_i$.  In our case the variables $x_i$ are $M$ and
the conserved charges $\tilde{q}_i$.  When $S$ is a smooth function of the
$x_i$, subadditivity is equivalent to the Hessian matrix $[{{\partial^2
S}\over{ \partial x_i \partial x_j}}]$ being negative definite.  Again
provided $S$ is smooth, the region of local stability is bounded by a locus
of simple zeroes of the determinant of the Hessian (or possibly zeroes of
higher odd order---but this is a non-generic situation that we will never
encounter).

On can also characterize the region ${\cal S}$ of phase space in which
local thermodynamic stability holds in terms of Legendre transforms.
${\cal S}$ is the largest possible region satisfying two criteria.  First,
it contains a particular reference point which we know is stable: for
instance, a point where $M$ is very large and all $\tilde{q}_i = 0$.
Second, in Legendre transforming the thermodynamic potential with respect
to some or all of the $x_i$, the map from $x_i$ to the thermodynamic
conjugate variables $\tilde{x}_i$ must be one-to-one, so that the Legendre
transform with respect to the chosen $x_i$ composed with the Legendre
transform with respect to the conjugate $\tilde{x}_i$ gives the identity
(i.e., the Legendre transforms are invertible).  The region ${\cal S}$ can be
parametrized in terms of $x_i$ or $\tilde{x}_i$.  Thus we see that
thermodynamic stability does not depend on the choice of ensemble; rather,
it is a criterion for being able to Legendre transform freely among
ensembles.  From a mathematical point of view, different ensembles are
different methods of calculating the same thing.

There is one big caveat to the foregoing remarks: we have to decide which
quantities are thermodynamic variables which could in principle be allowed
to vary in an experiment, and which are fixed parameters of the system
which cannot vary.  Striking out entries in the list of thermodynamic
variables $x_i$ actually decreases the dimension of the phase space.  For
example, if we regarded all the charges $\tilde{q}_i$ as fixed parameters,
then the phase space would be parametrized just by the mass $M$.  However,
in practice it is convenient to speak loosely of ``phase space'' as the
space of possible values for all the $x_i$, regardless of which of them we
regard as thermodynamic variables.  The more $x_i$ we regard as fixed
parameters, the larger the region of stability.  The reason is that we are
decreasing the family of Legendre transforms that have to be invertible.
Fewer conditions means a larger region on which they are all satisfied.

As a form of rhetorical shorthand, we shall indicate which of the $x_i$ we
regard as thermodynamic variables by specifying a particular ensemble.  For
instance, the canonical ensemble is the one obtained from the
microcanonical ensemble by trading entropy, or equivalently mass, for
temperature via a Legendre transform.  Thus any reference to the canonical
ensemble really means that we are thinking of the mass as the only
thermodynamic variable in the microcanonical ensemble, and that the charges
$\tilde{q}_i$ are fixed parameters.  The grand-canonical ensemble is
obtained from the microcanonical ensemble by Legendre transforming with
respect to entropy and all the charges, so we are now thinking of the
charges as thermodynamic variables, too.

The local stability computation in any ensemble---where by speaking of one
ensemble or another we are distinguishing thermodynamic variables from
fixed parameters, as explained in the previous paragraph---can in principle
be carried out by finding zeroes of the determinant of a submatrix of
$[{{\partial^2 S}\over{\partial x_i \partial x_j}}]$.  The appropriate
submatrix is the one which includes only those rows and columns pertaining
to the $x_i$ which are thermodynamic variables.  For the canonical
ensemble, this determinant is essentially the specific heat at constant
$\tilde{q}_i$.  For the grand-canonical ensemble, it is a combination of
this specific heat with other susceptibilities.  Direct computation of
these quantities tend to be extremely burdensome.  We find it more
efficient to work with the thermodynamic potential of the specified
ensemble, but as a function of the extensive thermodynamic variables rather
than the thermodynamic variables which are usually regarded as the
independent variables for that ensemble.  The method is best illustrated by
an example.  Suppose we are working in the canonical ensemble, so the
$\tilde{q}_i$ will always be regarded as fixed (and from here on
notationally suppressed).  Consider the Helmholtz (Euclidean) action:
  \eqn{IHDef}{
   I_H = \beta F = \beta M - S \ .
  }
 Suppose we write $M$ explicitly in terms of $S$: this amounts to a
complete specification of the thermodynamics in the microcanonical
ensemble.  Thus we write $I_H$ as a function of $S$ and the free parameter
$\beta$, which will be determined by the Legendre transform condition:
  \eqn{LCondition}{
   \left( {\partial I_H \over \partial S} \right)_\beta 
     = \beta {\partial M \over \partial S} - 1 = 0 \ .
  }
 If we take a second derivative, 
  \eqn{LSecond}{
   \left( {\partial^2 I_H \over \partial S^2} \right)_\beta
    = \beta {\partial^2 M \over \partial S^2} 
    = -{\partial^2 S/\partial M^2 \over (\partial S/\partial M)^2} \ ,
  }
 (where we have used \LCondition\ in the second equality), then we see that
$\left( {\partial^2 I_H \over \partial S^2} \right)_\beta = 0$ is
equivalent to ${\partial^2 S \over \partial M^2} = 0$, which is the
relevant local thermodynamic stability equation.  In fact, \LSecond\ is
still rather tedious to compute, but a change of variables makes life much
easier: we find we can write $S$ wholly in terms of $r_+$, which is the
location of the outer horizon (see equation \eno{SDef}).  It does not
matter that the $\tilde{q}_i$ enter into the relation between $S$ and
$r_+$, since they are just fixed parameters.  Next we note that the
condition
  \eqn{LThird}{
   \left( {\partial^2 I_H \over \partial r_+^2} \right)_\beta = 
     \left( {\partial S \over \partial r_+} \right)^2 
     \left( {\partial^2 I_H \over \partial S^2} \right)_\beta + 
     {\partial^2 S \over \partial r_+^2} 
     \left( {\partial I_H \over \partial S} \right)_\beta = 0
  }
 in the presence of the constraint \LCondition\ is equivalent to the
vanishing of \LSecond, provided ${\partial S / \partial r_+}$ is
non-singular.  Equation \LThird\ represents the essence of the method we
use to evaluate local thermodynamic stability.  Working in the
grand-canonical ensemble, there are more thermodynamic variables
($\tilde{q}_i$ as well as the mass or the entropy), and the convenient
reparametrization is in terms of $r_+$ and $q_i$.  Working through the
steps \LCondition, \LSecond, and \LThird, it is straightforward to show
that the zeroes of the determinant of the Hessian of $S$
with respect to $M$ and the $\tilde{q}_i$ coincide with the zeroes of the
determinant of the Hessian of the  Gibbs  (Euclidean) action $I_G$, written in the form
\begin{equation}
I_G=\beta(M-\sum_{i=1}^3{\tilde q}_i\phi_i)-S \ ,
\label{gibbsaFirst}
\end{equation}
 with respect to $r_+$ and $q_i$.  Where ${\partial r_+ \over \partial S}$
appears in \LThird, it will be replaced in the grand-canonical analysis by
the Jacobian ${\cal J} = \det {\partial (S,\tilde{q}_i) \over \partial
(r_+,q_i)}$.  It is straightforward to check that this Jacobian is
non-singular.  This completes our discussion of local thermodynamic
stability.

The Euclidean version of an R-charged black hole in $AdS_5$ is
characterized by some boundary data: the period for Euclidean time, and the
potential, which in ten-dimensional language amounts to the angular
velocity with which the $S^5$ is rotating.  There is another geometry with
the same boundary data: perfect, Weyl-flat, Euclidean $AdS_5$ times an
$S^5$ which again is rotating with the angular velocity specified by the
potential.  In five-dimensional terms, this geometry is interpreted as a
gas of R-charged particles in anti-de Sitter space.  Following Hawking and
Page \cite{HP}, we can ask whether this geometry or the black hole geometry
is favored.  The geometry with smaller action will dominate the Euclidean
path integral.  The nature of the boundary data determine how the action
must be computed.  If it is indeed the potential and the temperature which
are specified at the boundary, then the relevant action is the Gibbs
action, \eno{gibbsa}.  If instead the total R-charge is fixed, it is
appropriate to compute the Helmholtz action, \IHDef.  We will discuss the
computation of these actions in more detail in the next two Subsections.
In both cases our strategy will be to subtract off the pure $AdS_5$ action
from the black hole action to obtain a finite quantity, $I_G$ or $I_H$.
The vanishing of these actions will indicate the Hawking-Page phase
transition.  Past this transition point the black hole cannot be the global
minimum of the Euclidean action.  If the pure anti-de Sitter geometry is
the only geometry which competes with the black hole in the path integral,
and if the black hole is locally stable, then the Hawking-Page transition
represents the edge of the region where the black hole is globally stable.

To summarize, in judging local stability the choice of ensemble amounts to
a decision regarding which quantities in the system are thermodynamic
variables which are allowed to vary within the system, and with respect to
which Legendre transforms should be invertible; whereas in locating the
Hawking-Page transition, the choice of ensemble is motivated by the more
standard notion of whether a given thermodynamic quantity or its conjugate
(in the sense of Legendre transforms) is controlled from outside.  

It is a matter of physical motivation which ensemble one decides to
use.\footnote{In the context of spinning branes the distinction between the
canonical and grand-canonical ensembles and the correspondingly different
local stability constraints were addressed by Cai and Soh \cite{Cai}.}  For
instance, in the study of spinning branes \cite{CvGu} where the
world-volume is infinite, any given part of the brane can be regarded as a
``system'' in the presence of a ``heat bath'' which is the rest of the
brane.  The boundary between the system and the heat bath is fictitious,
and both energy and charge can flow across it.  This suggests that for the
purpose of judging local stability, one had better use the grand canonical
ensemble.  Using the canonical ensemble, where Legendre transforms with
respect to the charge density are disallowed, one is ignoring the
possibility that the charge density might become inhomogeneous.  

Holography also seems to favor the the grand-canonical ensemble, since the
temperature and electric potentials (which are fixed in the
grand-canonical ensemble) can be specified in terms of the asymptotic
geometry.  Specifically, the inverse temperature is the circumference of
the Euclidean time's $S^1$, and the electric potentials divided by the
temperature are the twist one makes before identifying around this circle.
The standard AdS/CFT approach would be to sum over all geometries with the
given asymptotic circumference and twist.  Assuming that only the black
hole geometry and the pure anti-de Sitter geometry are competing in the
sum, one judges which is preferred according to which has a lower Gibbs
free energy.  From this point of view the grand-canonical ensemble is the
relevant one, and we study it in Subsection~\ref{Gibbs} by computing the
Gibbs Euclidean action.  Partly to illustrate some interesting differences
between the ensembles, we calculate the Helmholtz Euclidean action in
Subsection~\ref{Helm} and discuss the stability constraints that follow
from it.

\subsection{Grand-Canonical Ensemble} 
\label{Gibbs}

In this subsection we spell out the properties of the Euclidean effective
action for the grand-canonical ensemble of R-charged black holes discussed
in Section~\ref{solution}.  A state in the grand-canonical ensemble is
usually specified by the inverse temperature $\beta$ and the electric
potentials $A_t^i \equiv \phi_i$ ($i=1,2,3$). 
% The heat and the charges
%$\tilde{q}_i$ are allowed to flow in and out of the system in order to keep
%$\beta$ and $\phi_i$ fixed.\fixit{Do we really want to say this?}

The Gibbs Euclidean action can be obtained following the procedure,
initiated by Hawking and Page~\cite{HP} for the Schwarzschild black hole in
$AdS_4$, further developed by York {\it et al.}~\cite{Yorketal}, and
applied to $D=4$ anti-de Sitter Reissner Nordstr\"om black holes in
\cite{Louko} (see also \cite{Peca}).  When employing this procedure the
boundary conditions (including the local temperature) are defined at a
finite value of the radial coordinate $r_B$; these boundary conditions in
turn uniquely determine the reduced Euclidean action which has both bulk
and surface term contributions.  One also subtracts from it the pure
$AdS_5$ space-time contribution.  As the last step the limit $r_B\to
\infty$ is taken, i.e., the artificial boundary is removed.
For further details see, e.g., \cite{Yorketal,Peca,Louko}.\footnote{In
Ref.~\cite{witHolTwo,GubsKleTs} an alternative approach in the study of the
of AdS-Schwarzschild black holes, suitable for the study of the higher loop
($\alpha'$) corrections to the effective action, was developed. There one
considers only a bulk term, but again there is a cutoff at large radius
both for the pure $AdS$ geometry side and for the AdS-Schwarzschild black
hole side.  The pure $AdS$ geometry is periodized with a temperature in
such a way that at the cut-off radius the circumference of the compactified
$S^1$ is the same as for AdS-Schwarzschild black hole.  Then by subtracting
the pure $AdS$ bulk integral from the AdS-Schwarzschild black hole bulk
integral, one obtains a finite result corresponding to the
AdS-Schwarzschild Euclidean action.}

The above procedure will assign zero action to the pure $AdS_5$ geometry.
With no subtractions and no cutoff, the action integral formally diverges,
so some choice of zero point is necessary.  The $AdS_5$ geometry also
involves a constant electric potential.  Constant electric potential yields
zero field strength and hence makes no contribution to the action, and the
the anti-de Sitter geometry is undistorted.  In the ten-dimensional
geometry, the constant electric potential is interpreted as a twist on the
$S^5$ part of the geometry.  To understand this, note that the gauge fields
arise from mixed components of the ten-dimensional metric, $g_{\mu\alpha}$,
where $\mu$ is an $AdS_5$ index and $\alpha$ is an $S^5$ index.  For
instance, if we turn on a single constant electric potential, corresponding
to rotations in the $\phi$ direction of an $S^5$ parametrized so that the
metric is $ds^2 = d\theta^2 + \sin^2\theta d\phi^2 + \cos^2\theta
d\Omega_3^2$, then what we are doing from a ten-dimensional perspective is
making $g_{t\phi}$ non-zero.  This off-diagonal entry in the metric can be
exactly killed if one changes the angular variable from $\phi$ to
$\tilde\phi = \phi - \Omega t$.  $\Omega$ has the interpretation of an
angular velocity at which the $S^5$ rotates.  Now, $\tilde\phi$ is single
valued in the periodized Euclidean geometry, whereas $\phi$ jumps by
$\Omega/T_H$ at the identification point.  This is what we mean by twist on
the $S^5$.  If one applies Kaluza-Klein reduction to get back to
five-dimensions, then in the normalization conventions used in
Section~\ref{equiv}, we have the electric potential $A_t^{\tilde\phi} =
-\Omega$.

The Gibbs Euclidean action takes the following form:

\begin{equation}
I_G=\beta(M-\sum_{i=1}^3{\tilde q}_i\phi_i)-S
\label{gibbsa}
\end{equation}
 where $M$ is the ADM mass, ${\tilde q}_i$ is the physical charge and $S$
is the entropy.  $S={A\over {4G_5}}$ where $A$ is the area at the outer
horizon.

The structure of this action is expected from thermodynamic considerations
as well.  The physical, Lorentzian, black hole solutions are extrema of
(\ref{gibbsa}).  The expressions \eno{beta} and \eno{phi} for the inverse
Hawking temperature $\beta$ and the electric potentials $\phi_i$ at the
outer horizon can be obtained by extremizing \eno{gibbsa} with respect to
$S$ and $\tilde{q}_i$ with $\phi_i$ and $\beta$ held fixed.  As discussed
at the beginning of this Section, these extrema can also be obtained by
finding zeroes of the first derivatives of (\ref{gibbsa}) with respect to
$q_i$ and $r_+$, again keeping $\beta$ and $\phi_i$ fixed, provided the
Jacobian ${\cal J} = \det {\partial (S,\tilde{q}_i) \over \partial
(r_+,q_i)}$ is non-singular.

For the R-charged black holes of Section \ref{solution} the Gibbs action
(\ref{gibbsa}) can be expressed explicitly in
  terms of $q_i$ and $r_+$, and it takes the 
 following suggestive form:
 \begin{equation}
  I_G^{BH}={\beta\over{2r_+^2}}\left[r_+^4-\prod_{i=1}^3(r_+^2+q_i)\right] \ ,
 \label{gibbsabh}
  \end{equation}
 where $\beta$ is determined by (\ref{beta}).  (We have set $L=1$, but
powers of $L$ can be restored using the replacements described at the end
of section~\ref{solution}).

Now let us specialize to only one non-zero charge: say $q_1= q\ne 0$ but
$q_2 = q_3 = 0$.  The local thermodynamic stability is most efficiently
analyzed by computing the Hessian of \eno{gibbsa} with respect to $r_+$ and
$q$ with $\beta$ and $\phi$ held fixed, in analogy with \LThird.  The
stability constraint is 
\begin{equation}
2r_+^4+r_+^2(q+1)-(q-1)^2\ge 0 \ . 
\label{grandcanstab}\end{equation}
 In the phase space parametrized by $(M,\tilde{q})$, or more conveniently
by $(r_+,q)$, the zeroes of the left hand side of \eno{grandcanstab}
determine the critical lines which form the boundary of stability.  There
are two branches:
 \begin{equation}
q_{\pm}(r_+)=1+\textstyle{1\over 2}r_+^2\pm\textstyle{1\over 2}
\sqrt{8r_+^2+9r_+^4}\ .
\end{equation}
 The regions for which $q_{-}(r_+)\le q \le q_{+} (r_+)$ correspond to
locally stable regions, i.e., $I_G^{BH}$ is a local minimum.  When $q_-=0$
the critical point is at $r_+={1\over\sqrt{2}}$ [${\tilde q}=0$, $M={9\over
8}$]~\cite{BCS}.  The critical lines merge at $q_+=q_-=1$ and $r_+=0$
[$M={\tilde q}=1$]. This point is a BPS solution with zero entropy and
finite $\beta=\pi$.  Black holes in the regions $q<q_-(r_+)$ and
$q>q_+(r_+)$ are thermodynamically unstable. The region $q<q_-(r_+)$
charged black holes which are much smaller than the size of anti-de Sitter
space: $r_+\ll 1$, $q\ll 1$.  The curvature of $AdS$ is negligible in the
vicinity of these black holes, so we are predicting that singly charged
black holes should be unstable even in flat space.  And they are: they can
be obtained in a $T^5$ compactification of type IIB string theory as
NS5-branes or D5-branes wrapped around the $T^5$.  The thermodynamic
instability of NS5-branes is familiar: the temperature in the limit of
extremality is the Hagedorn temperature of the fractionated instanton
strings on the world-volume \cite{juanFive}, and as non-extremality
increases the temperature goes down.

The location of the Hawking-Page phase transition between the black hole
solution and the pure $AdS_5$ solution is determined by the constraint
$I_G^{BH}=0$:
 \begin{equation}
 q_0(r_+)=1-r_+^2 \ ,
 \end{equation}
 where we have used \eno{gibbsabh}.  For $q>q_0(r_+)$ the black hole
solution has $I_G^{BH}<0$ and thus it is the dynamically preferred
solution.  $q_0(r_+)=0$ corresponds to $r_+=1$ [${\tilde q}=0$, $M=3$].
This is the Hawking-Page phase transition for the $D=5$ $AdS$-Schwarzschild
black hole solution.
 
For $q>q_+(r_+)>q_0 (r_+)$, black holes are favored over the pure anti-de
Sitter space-time, but they are locally unstable.  In this case our
evaluation is that we have failed to find the entropically preferred state
of the system.  One possibility that suggests itself is that a multi-black
hole solution might be favored, but it seems difficult to imagine a static
solution describing multiple black holes in $AdS_5$.  

For $q_-(r_+)< q<q_0(r_+)$, pure anti-de Sitter space is entropically
favored over the black hole solution, but the black hole solution still
corresponds to a local minimum of the Gibbs action.  In the uncharged case,
this is the region between $r_+={1\over \sqrt{2}}$ [$M={9\over 8}$] and
$r_+=1$ [$M=3$].  These black holes could be termed ``meta-stable'' in the
sense that they can exist in equilibrium with a heat bath of particles in
anti-de Sitter space and are stable under small perturbations, but the
system as a whole could gain entropy by converting the entire black hole
into a gas of R-charged particles.
 
In Figure~1a we show all these stability domains in a plot of $q$
vs.~$r_+$.  To help the reader's intuition we also exhibit the same
stability plots in Figure~1b, but now as ${\tilde q}$ vs.~$M$.  (On this
graph only the region $M\ge {\tilde q}$ is consistent with the BPS bound.)
The vertically shaded areas correspond to the regions where the $AdS_5$ is
the preferred solution, and the horizontally shaded area correspond to the
regions where the black hole solutions are local minimum of the Gibbs
action.

\begin{figure}
   \vskip0cm
   \centerline{\psfig{figure=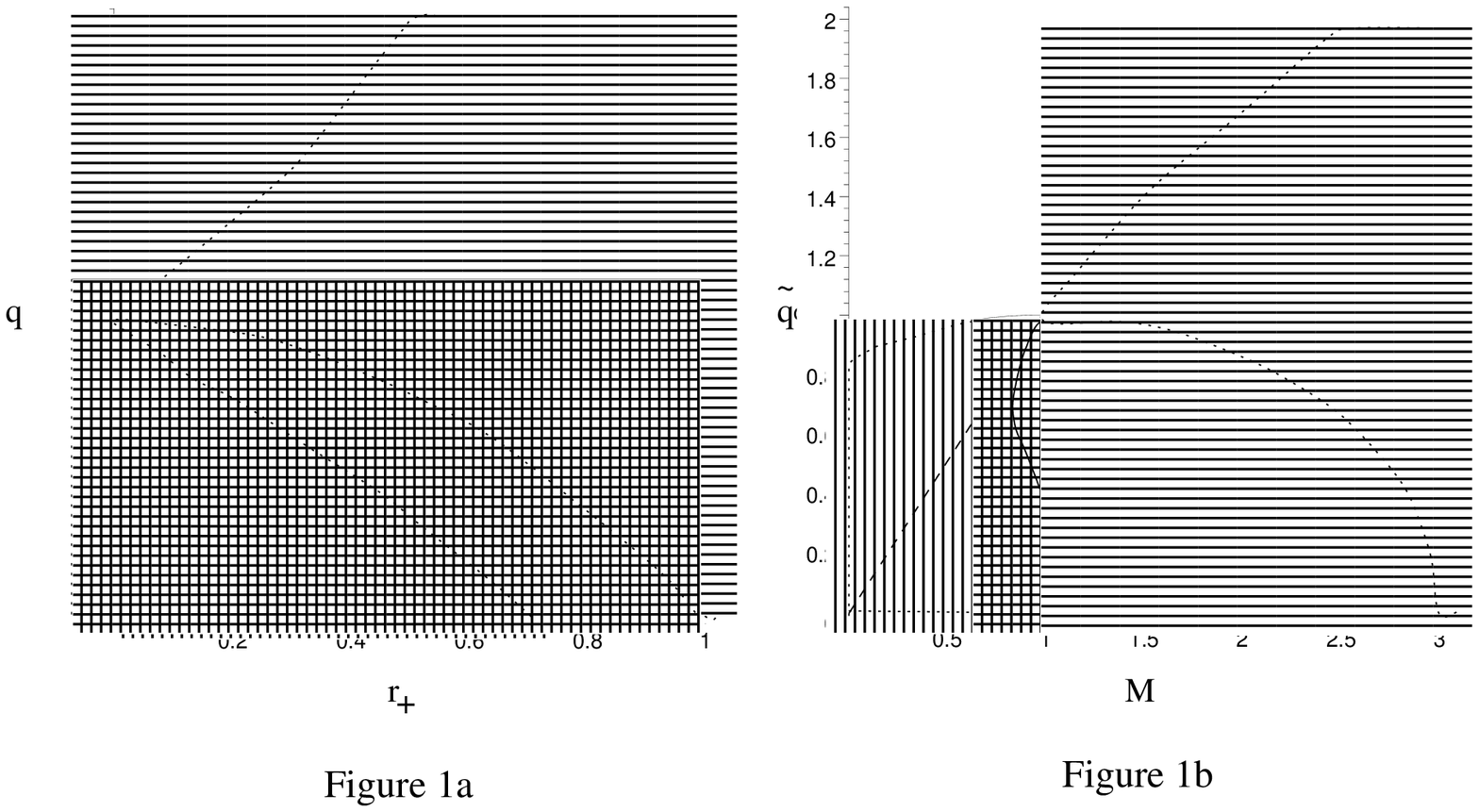,width=5.5in}}
   \vskip1cm
 \caption{In Figure~1a stability domains are plotted as ${q}$ vs.~${{r_+}}$
for the grand-canonical ensemble of $D=5$ R-charged black holes with one
charge turned on.  In Figure~1b the same stability plots are exhibited, but
now as ${\tilde q}$ (the physical charge) vs.~$M$ (the ADM mass).  Only the
region satisfying the BPS bound, $M\ge {\tilde q}$, corresponds to physical
black hole solutions.  The vertically shaded areas correspond to the
regions where the $AdS_5$ is the preferred solution and the horizontally
shaded area correspond to the regions where the black hole solutions are
local minimum of the Gibbs action.  The checkered area is the domain of
common overlap: there black holes still correspond to the local maxima of
the entropy (local minima of the Euclidean action); however, anti-de Sitter
space is preferred globally.  In the unshaded area, neither the black
hole solution nor pure anti-de Sitter space are global minima of the
action.}\label{figure1}
  \end{figure}

\subsection{Canonical Ensemble}
\label{Helm}

We now repeat the stability analysis for the canonical ensemble; along with
the fixed inverse temperature $\beta$ , one now fixes the physical charges
${\tilde q}_i$, i.e., the field strengths $F_{\mu,\nu}^i$ (and not the
electric potentials $\phi_i$) are fixed.  Following a procedure analogous
to the one spelled out for the grand-canonical ensemble in
Subsection~\ref{Gibbs}, but now with the new boundary conditions, one
obtains the following form of the Helmholtz Euclidean action:
 \begin{equation}
I_H=\beta M-S\ , 
\label{helma}
\end{equation}
 where $M$ and $S$ are the mass and the entropy of the system and $\beta$
is the inverse temperature.  Extremizing the above action with respect to
$S$, or equivalently, with respect to $r_+$ (while keeping the physical
charges ${\tilde q}_i$ fixed) determines the correct inverse temperature
$\beta$ of the black hole solution (\ref{beta}).
 
To locate the Hawking-Page transition, we write the Helmholtz action
(\ref{helma}) for the black hole solution described in
Section~\ref{solution} as a function of $q_i$ and $r_+$:
 \begin{equation}
  I_H^{BH}={\beta\over{2r_+^2}}\left[r_+^4(1-r_+^2)+r_+^2\sum_{i=1}^3q_i(2+r_+^2)
  +3r_+^2\sum_{i<j}^3q_iq_j+5\prod_{i=1}^3q_i\right]\ , 
 \label{helmabh}
  \end{equation}
 where $\beta$ is defined in (\ref{beta}).  The phase transition between
the black hole solution and the pure $AdS_5$ solution is determined by the
zero of $I_H^{BH}$ (\ref{helmabh}):
 \begin{equation}
 q_0(r_+)={{r_+^2(r_+^2-1)}\over {2+r_+^2}} \ .  
 \end{equation}
 In the domain $q>q_0(r_+)$ the black hole solution has $I_H<0$ and
thus it is the dynamically preferred solution. (For $q_0=0$ the critical
point is again $r_+=1$ [${\tilde q}=0$, $M=3$].)  

To evaluate local thermodynamic stability, we use the method outlined in
Subsection~\ref{LocalGlobal} around equation~\LThird.  In the case of a
single non-zero charge, say $q_1=q\ne 0$, the local stability constraint
leads to:
 \begin{equation}
2r_+^4+r_+^2(5q+1)-q^2+6q-1\ge 0 \ . 
\label{canstab}\end{equation}
 This case is simple enough that one can easily verify \eno{canstab} by
explicitly computing $\left( {\partial^2 S \over \partial M^2} \right)_q$.
 The zeroes of the left hand side of \eno{canstab} form two critical lines:
\begin{equation} q_{\pm}(r_+)=1+\textstyle{5\over
2}r_+^2\pm\textstyle{1\over 2} \sqrt{32+ 64r_+^2+33r_+^4}\ , \end{equation}
 and the stable region corresponds to the domain
  $q_{-}(r_+)\le q\le q_{+}(r_+)$. 
 Again, $q_-=0$  corresponds to the critical point:
  $r_+\le {1\over\sqrt{2}}$ [${\tilde q}=0$, $M={9\over 8}$]~\cite{BCS}. 
 On the other hand now the stable domain with $r_+=0$ corresponds to the
range of charges:
  $3-2\sqrt{2}\le q\le 3+2\sqrt{2}$. 
 [Equivalently, this is the region with  $M={\tilde q}$ (BPS-limit) and 
  $3-2\sqrt{2}<{\tilde q}=q< 3+2\sqrt{2}$.]  
 Black holes in the region $q<q_-(r_+)$ and $q>q_+(r_+)$ are not
thermodynamically stable.
 
In Figure~2a we show the stability regions in a plot of $q$ vs.~$r_+$.  In
Figure~2b we exhibit the same stability regions plotting ${\tilde q}$
vs.~$M$.  The vertically shaded areas correspond to the regions where the
$AdS_5$ is the preferred solution and the horizontally to the regions where
the black hole solution is the local minimum of the Helmholtz action.
\begin{figure}
   \vskip0cm
   \centerline{\psfig{figure=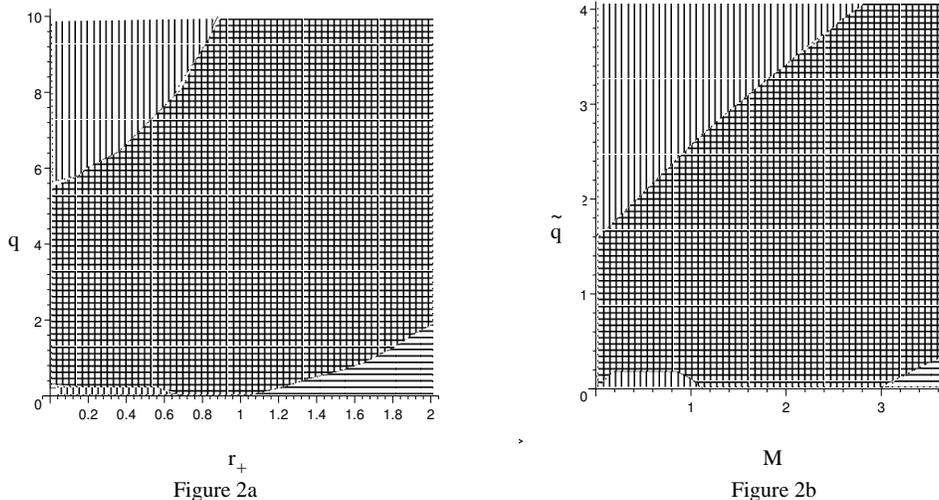,width=5in}}
   \vskip0cm
 \caption{In Figure~2a we plot the stability domains as $q$ vs.~$r_+$ for
the canonical ensemble of $D=5$ R-charged black holes with one non-zero
charge.  In Figure~2b the same stability plots are exhibited, but now
plotting ${{\tilde q}}$ vs.~${M}$.  The vertically shaded areas are the
regions where anti-de Sitter space is the preferred solution.  The
horizontally shaded areas are the regions where the black hole solution is
a local minimum of the Helmholtz action. The checkered area is the domain of
common overlap: meta-stable black holes in the intepretation of
section~\ref{Gibbs}.}\label{figure2}
  \end{figure}

Note three salient differences between the results for the canonical and
grand-canonical ensembles.  First, as expected from the general
considerations of Subsection~\ref{LocalGlobal}, the region of local
stability for the canonical ensemble is larger than for the grand-canonical
ensemble.  Second, the region favored by the pure $AdS_5$ is significantly
increased for the canonical ensemble, and it encompasses the whole region
with $r_+<1$ and non-zero $q_i$'s.  Third, and most interesting, the pure
$AdS_5$ space-time is dynamically favored in the whole domain of the black
hole parameters for which the black holes are locally unstable.

\section{Spinning D3-branes as R-charged black holes}
\label{RCharge}

In this Section we establish the precise connection between the
near-extreme spinning D3-branes solution and the solutions of $N=8$, $D=5$
gauged supergravity found in \cite{BCS} and discussed in
Section~\ref{solution}.  In this Section we restore all factors of $L$
explicitly, but still set $G_5 = {\pi \over 4}$.

\subsection{Equivalence of $k=0$ R-charged  black holes and near-extreme
spinning D3-branes}
\label{k0}

Angular momenta in planes perpendicular to the world-volume of D3-branes
(which we refer to as spins to distinguish them from angular momenta in
other planes) are realized in the world-volume theory as charges under the
global $SO(6)$ R-symmetry group.  In the reduction of type~IIB supergravity
on $S^5$, the $SO(6)$ isometry becomes the non-abelian gauge symmetry of
the resulting ${\cal N}=8$ gauged supergravity in five dimensions, and the
spins become $SO(6)$ gauge charges.  In the AdS/CFT correspondence, the
$SO(6)$ supergravity gauge fields in the bulk of $AdS_5$ couple to the
gauge theory R-currents on the boundary.  It is well-known that the
near-horizon geometry of near-extreme D3-branes without spin reduces to a
Schwarzschild black hole in $AdS_5$.  The purpose of this section is to
show how spin on the D3-branes reduces to charge on the black holes.  By
explicitly performing the Kaluza-Klein reduction on $S^5$, we will obtain
the precise relationship between the parameters of the R-charged black
holes and those of the spinning D3-branes.  For the sake of simplicity only
one angular momentum, so that the black hole will have $q_1=q\ne 0$ but
$q_2 = q_3 = 0$.  One can verify that the computation extends to the case
of all three angular momenta non-zero.

The near-horizon form of the ten-dimensional D3-brane solution with a single
angular momentum is~\cite{CvYou2,Larsen}
  \eqn{TenSol}{\eqalign{
   ds_{10}^2 &\equiv \hat{g}_{MN} dx^M dx^N  \cr
    &= {1 \over \sqrt{f}} \Bigg[ -h dt^2 - 
     {2 \ell L^2 r_0^2 \over r^4 \Delta} \sin^2 \theta dt d\phi + 
     f \tilde\Delta r^2 \sin^2 \theta d\phi^2  \cr
    &\qquad{} + {f \over \tilde{h}} dr^2 + 
     d\vec{x}^2 + f\Delta r^2 d\theta^2 + 
     f r^2 \cos^2 \theta d\Omega_3^2 \Bigg]  \cr
   f &= {L^4 \over r^4 \Delta}  \qquad
   \Delta = 1 + {\ell^2 \cos^2 \theta \over r^2}  \qquad
   \tilde\Delta = 1 + {\ell^2 \over r^2}  \cr
   h &= 1 - {r_0^4 \over r^4 \Delta}  \qquad
   \tilde{h} = {1 \over \Delta} \left( 1 + {\ell^2 \over r^2} - 
    {r_0^4 \over r^4} \right) \ .
  }}
 The basic Kaluza-Klein ansatz is \cite{mahsch}
  \eqn{KKAnsatz}{
   \hat{g}_{MN} = \pmatrix{ e^{-{4 \over 3} D} g_{\mu\nu} + 
     A_\mu{}^\gamma A_{\mu\gamma} & A_{\mu\beta}  \cr
     A_{\nu\alpha} & G_{\alpha\beta}} \ ,
  }
 where indices $\alpha,\beta,\gamma$ run over the compact internal manifold
$K$, and $\mu,\nu$ run over the non-compact spacetime $M$.  The metric on
$K$ can vary with the position on $M$, and the equation
  \eqn{DilDef}{
   \vol_K = e^{2D} \vol_{K,\infty} 
  }
 defines the ``dilaton'' field $D$ (not to be confused with the
ten-dimensional dilaton $\phi$) in terms of how the volume form on $K$
compares to its limit at spatial infinity.  The power of $e^D$ in
\KKAnsatz\ is chosen to put the metric $g_{\mu\nu}$ in Einstein frame.  For
the solution \TenSol\ we have 
  \eqn{KMet}{\eqalign{
   ds_K^2 &\equiv G_{\alpha\beta} dx^\alpha dx^\beta = 
    \sqrt{f} \left[ \Delta r^2 d\theta^2 + 
    \tilde\Delta r^2 \sin^2 \theta d\phi^2 + 
    r^2 \cos^2 \theta d\Omega_3^2 \right]  \cr
   e^{2D} &= {\sqrt{\tilde{\Delta}} \over \Delta^{3/4}} \ .
  }}
 The only non-zero component of $A_\mu{}^\gamma$ is 
  \eqn{NonzeroA}{
   A_t{}^\phi = -{\ell r_0^2 \over L^2 r^2 \tilde\Delta} \ .
  }
 Finally, the five-dimensional metric $g_{\mu\nu}$ is just the solution
(\ref{metric})  with $k=0$, $\mu = r_0^4/L^2$, $q_1 = \ell^2$, and $q_2 = q_3 = 0$.
Note that although $D$ and $ds_K^2$ vary over both $M$ and $K$, the
five-dimensional metric $ds_M^2$ and the five-dimensional gauge potential
$A_t{}^\phi$ are independent of the internal coordinates.  The dependence
of $D$ on $K$ indicates that the compactification is a warped product.

There is a slight subtlety in comparing the gauge fields in (42) of
\cite{BCS} with \NonzeroA: superficially in \cite{BCS} the field strength
seems to vanish because of an explicit prefactor of $\sqrt{k}$.  At this
point it is useful to recall the approach of \cite{witHolTwo} where the
$k=0$ solution is recovered by focusing in on a small, nearly flat part of
the $S^3$ in the $k=1$ solution.  This amounts to taking the limit $k \to
0+$.  In this limit, $q = \ell^2$ and $\mu = r_0^4/L^2$ are held fixed.
From equation (43) of \cite{BCS} one obtains $\sinh^2 \beta = k \ell^2
L^2/r_0^4$.  The factor of $k$ in this relation conspires to cancel the one
explicitly shown in (42) of \cite{BCS}, and the end result for the gauge
field $A^1$ pertaining to the charge $q_1$ is
  \eqn{AOneFinal}{
   A^1_t = {L \over 2} A_t{}^\phi \ .
  }
 The factor of ${L\over 2}$ comes from a different choice of normalization of
$A_\mu$ in \cite{BCS}.  For us the gauge kinetic term picks up a factor of
$L$ from $\sqrt{G} G^{-1} G^{-1}$.

There is some debate over the status of five-dimensional ${\cal N} = 8$
gauged supergravity as a truncation of the Kaluza-Klein reduction of type
IIB supergravity/string theory.  In particular, is it true that solutions
of the former can always be promoted to solutions of the latter?  This
seems guaranteed for supersymmetric solutions from closure properties of
the supersymmetry algebra, but for non-supersymmetric solutions one is
making a nontrivial statement regarding the equations of
motion.\footnote{We thank J.~Distler for explaining this to us.}  Modulo
the small technical issue of compactifying ${\bf R}^3$ to $S^3$,
non-extreme R-charged black holes provide an example of a
non-supersymmetric solution of gauged supergravity which can indeed be
promoted to a solution in ten dimensions: namely the spinning D3-brane
solution.  The key point is that the gauge potential $A_t{}^\phi$ and the
metric $g_{\mu\nu}$ resulting from the Kaluza-Klein compactification of the
spinning D3-brane solution are independent of the $S^5$ coordinates.

\subsection{Thermodynamic equivalence of large R-charged black holes and
 spinning D3-branes}
\label{equiv}

Another way to view the $k\to 0^+$ limit is that black hole solutions with
$k=0$ can be recovered from large black hole solutions with $k=1$ by
restricting attention to a angular region of the $S^3$ in the $k=1$ region
where the curvature can be neglected.  This was explained precisely for
Schwarzschild black holes in \cite{witHolTwo}, and the same story goes
through with charged black holes.  New physics shows up in the $k=1$
solutions (for instance, the confinement-deconfinement transition) because
of the finite-size effects of the $S^3$.  When we take $k\to 0^+$, we are
making the volume of the $S^3$ very large compared to the cube of the
inverse temperature.  Neglecting all finite size effects should lead us
back to the thermodynamics of spinning D3-branes whose world-volume is flat
and infinite.  The purpose of this Subsection is to show this explicitly
and to establish the precise connection between the respective physical
parameters.

Large R-charged black holes are the ones with $r_+\gg L$ and $q_i\gg L^2$.
In this limit, the expressions for the ADM mass $M$ (\ref{mass}) and the
physical charges ${\tilde q}_i$ (\ref{pcharge}) reduce to the following
form:
 \begin{equation}
M\sim \textstyle{3\over 2}\mu, \ \   
{\tilde q}_i^2\sim q_i\mu\ , \ \ (i=1,2,3) \ , 
\label{AddEnt}\end{equation}
where $\mu$ is related to 
$r_+$ and  $q_i$'s  by $f(r_+)=0$ (see
eq.(\ref{f})). In the large charge limit this equation implies:
\begin{equation}
{1\over L^2}\prod_{i=1}^3(r_+^2+q_i)-\mu r_+^2\sim 0\ .
\label{horizon}\end{equation}
 To obtain \eno{horizon}, one can start with $f(r_+) r_+^4 = 0$ and drop
the term $kr_+^4$.  This makes sense in view of the equivalence of $k\to
0^+$ and large $k=1$ black holes.

 By introducing new variables
\begin{equation}
y_H\equiv \alpha^{-1}\sqrt{\mu}r_+, \ \ y_i \equiv \alpha^{-1}{{\tilde q}_i}  \ , 
(i=1,2,3)\ , 
\label{newdef}\end{equation}
where $\alpha= (\mu^3 L^2)^{1/ 4}$, one can rewrite  (\ref{horizon}) in the
following form:
\begin{equation}
 \prod_{i=1}^3(y_H^2+y_i^2)-y_H^2=0\ .
\label{normalhor} \end{equation} 
This is precisely the ``horizon'' equation  for spinning D3-branes  displayed in
Section II of Ref. \cite{CvGu}.

One can introduce the  energy density $e$, entropy density $s$ and 
 the  charge density $j_i$ in the following way:
\begin{eqnarray}
e&\equiv& {M\over{N^2 V}}\ ,\nonumber \\
 j_i&\equiv& {{{\tilde q}_i}\over {N^2 V}}=
 {s\over{2\pi}}{y_i\over y_H}\ ,
  \nonumber\\
s&\equiv& {S\over{N^2 V}}=  2\pi \gamma e^{3/ 4}y_H \ ,
\label{normalent}\end{eqnarray}
where
\begin{equation}
V=2\pi^2 L^3, \ \ N^2= 2L^3, \ \ \gamma={2\over {27\pi^2}}\ .
\label{vol}\end{equation}
 The thermodynamic variables introduced in (\ref{normalent}), together with
the definition of the volume $V$ and the flux $N$ (\ref{vol}) in terms of
$L$ (and $G_5$ if we hadn't set it equal to $\pi/4$) and also the
relationship between $y_H$ and $y_i$ (\ref{newdef}), provides a precise
translation of the thermodynamic variables of large black holes into those
of spinning D3-branes.  It may seem strange to assign a definite volume to
the $S^3$ on the boundary of $AdS_5$.  However, as commented on in
\cite{witHolOne} and \cite{hs}, a choice of radial coordinate is equivalent
to choosing a definite metric among those with a specified conformal
structure.  Our choice of radial coordinate was fixed in \eno{metric}.

Even in view of the above translation of variables, the equivalence between
the local thermodynamic stability constraints for spinning branes and
R-charged black holes is slightly non-trivial: equations~\eno{AddEnt} and
\eno{normalent} define a non-linear relation between $e$ and $j_i$ for the
spinning brane and $M$ and $\tilde{q}_i$ for the R-charged black hole.  To
see that the stability constraints must turn out the same, think of this
relation as a reparametrization of phase space, similar to the convenient
$(r_+,q_i)$ parametrization we used in earlier sections.  The argument in
Section~\ref{LocalGlobal} should guarantee that the stability region does
not depend on the reparametrization.  In the case of one non-zero charge,
the stability constraints for R-charged black holes for the grand-canonical
ensemble (\ref{grandcanstab}) and canonical ensemble (\ref{canstab}) reduce
to the following respective forms in the $r_+/L \to \infty$, $q/L^2 \to
\infty$ limit:
 \begin{eqnarray}
&2&r_+^2-q= {\mu\over \alpha^2} y_H^2(2-x^2)\ge 0 \qquad\hbox{grand-canonical} \\
&2&r_+^4+5r_+^2q-q^2 ={\mu\over \alpha^2} y_H^2(2+5x^2-x^4)\ge 0 
  \qquad\hbox{canonical.}
\end{eqnarray}
 Here $x\equiv {y\over y_H}$.  These stability constraints are in precise
agreement~\cite{Gubser,Cai} with those of near-extreme spinning D3-branes
with only one angular momentum turned on.

\section{Phases of R-charged black holes in $D=4$}
\label{FourDim}

So far we have focused our attention on black holes in $AdS_5$ because they
are dual to thermal states of ${\cal N}=4$ supersymmetric gauge theory in
3+1~dimensions.  Our analysis is extended in this section and the next to
black holes in maximally supersymmetric gauged supergravity in $D=4$ and
$D=7$.  In Section~\ref{Mbranes} we outline their relationship with
spinning M2- and M5-branes.  

Black holes in $D=4$ $N=8$ gauged supergravity have been studied in
Ref.~\cite{Duff}.  Their Einstein frame metric is of the following
form~\cite{Duff,BCSII}:

\begin{equation}
ds^2 = - {(\prod_{i=1}^4 H_i)^{-1/2}} f dt^2 + (\prod_{i=1}^4 H_i)^{1/2}
 	\bigg(f^{-1} {dr^2} + 
	r^2 d\Omega_{2,k} \bigg) \ ,
	\end{equation}
where	  \\ 
\begin{equation}
f = k - {\mu \over r} + g^2 r^2 \prod_{i=1}^4H_i \ , \ \ H_i=1+{{q_i}\over r}, \
(i=1,\cdots 4) \ . 
\label{f4}\end{equation}
Here we have chosen $G_4={1\over 4}$. (Again  we shall concentrate on
black hole solutions with $k=1$.) 
With this choice of $G_4$, $q_i$'s 
and   the physical charges ${\tilde q}_i$'s  are defined in (\ref{charges}) and
the ADM mass is of the following form:
\begin{equation}
M=2 \mu +\sum_{i=1}^4q_i=
2(r_++r_+^3g^2\prod_{i=1}^4H_i)+\sum_{i=1}^4q_i \ , 
\end{equation}
where in the second equality the
radial coordinate at the outer horizon $r_+$  is defined as the largest
non-negative zero of $f(r)$ (\ref{f4}). 

Again the parameterization of the thermodynamic quantities in terms of 
$q_i$ and $r_+$ (instead of  ${\tilde q}_i$ 
 and $M$) is most suitable:
\begin{equation}
S= {A\over{4G_4}}=4\pi \sqrt{\prod_{i=1}^4 (r_++q_i)}\ ,
\label{ent4}\end{equation}
\begin{equation}
\beta={1\over T_H}={{4\pi r_+\sqrt{\prod_{i=1}^4
(r_++q_i)}}\over{3r_+^4+2r_+^3\sum_{i=1}^4q_i+r_+^2(\sum_{i<j}^4q_iq_j+1)-\prod_{i=1}^4q_i}} .
\label{beta4}\end{equation}
The physical charges are then determined in terms of $q_i$  and $r_+$
in the following way:
\begin{equation}
{\tilde q}_i^2=q_i(r_++q_i)\left[1 +{1\over r_+}\prod_{j\ne
i}(r_++q_j)\right] \ ,
\end{equation}
and the electric potentials at the outer horizon take the following  form:
\begin{equation}
\phi_i=A^i_t(r_+)={{\tilde q_i}\over {r_+ +q_i}} \ , \ (i=1,\cdots, 4)\ . 
\label{pot4}\end{equation}

In the following, along with 
$G_4 = {1\over 4}$ we shall also take  $L\equiv {1\over g}= 1$.  (Note $L$ is
related to the asymptotic cosmological constant $\Lambda = -{3\over L^2}$.) However, 
$L$ can  be restored
by  replacing 
 $r_+$  by ${r_+\over L} $ and  $q_i$ by ${q_i\over L}$.

\subsection{Grand-Canonical Ensemble  and 
Stability Constraints}

The Gibbs Euclidean action is of the form:

\begin{equation}
I_G=\beta(M-\sum_{i=1}^4{\tilde q}_i\phi_i)-S\ , 
\label{gibbsa4}
\end{equation}
 where $M$ is the mass, ${\tilde q}_i$ is the physical charge and
$S={A\over {4G_4}}$ is the entropy.  The extrema of this action correspond
to the black hole solutions for which the inverse Hawking temperature and
the electric potentials at the outer horizon are given in (\ref{beta4}) and
(\ref{pot4}), respectively.

The critical hypersurfaces of thermodynamic stability can again be
determined by evaluating the zeroes of the determinant of the Hessian of
second derivatives of (\ref{gibbsa4}) with respect to $r_+$ and $q_i$.
 
The phase transition between the black hole solutions and pure $AdS_4$
occurs when (\ref{gibbsa4}) for the black hole solution ceases to be
negative.
  When expressed explicitly in
  terms of $q_i$ and $r_+$,  (\ref{gibbsa4}) 
   takes the following suggestive form:
 \begin{equation}
  I_G^{BH}={\beta\over{r_+}}\left[r_+^2-\prod_{i=1}^4(r_++q_i)\right] \ ,
 \label{gibbsabh4}
  \end{equation}
where $\beta$ is determined by (\ref{beta4}). 

For one non-zero charge only, say $q_1= q\ne 0$, the the region of local
thermodynamic stability is specified by
\begin{equation}
3r_+^4 + q r_+^3  -  2q^2 r_+^2  +   2r_+^2  +  3q r_+ - 1\ge 0\ . 
\label{grandcanstab4}\end{equation}
The zeroes of the above expression correspond to two critical
lines:
\begin{equation}
q_{\pm}(r_+)={3\over{4r_+}}+{r_+\over{4}}\pm
{1\over 4}\sqrt{{1\over
r_+^2}+22+25r_+^2}\ , 
\end{equation}
with the stable domain satisfying $q_{-}(r_+)<q<q_{+}(r_+)$. 
$q_-=0$  has a critical point at
 $r_+={1\over\sqrt{3}}$ \cite{HP}.  Note also that as $r_+\to 0$ the local
stability region is pushed to large $q$: $q_{\pm} \to \infty$ as $r_+ \to
0$.
 
 The  phase transition between the black hole solution and the pure 
 $AdS_4$ solution is  now determined by:
 \begin{equation}
 q_0(r_+)={1\over r_+}-r_+ \ .  
 \end{equation}
 For $q>q_0(r_+)$ the black holes  are  dynamically preferred over $AdS_4$.  As 
 $r_+\to 0$, the critical line  is pushed to $q_0\to \infty$.
 $q_0=0$  has the  Hawking-Page transition at  $r_+=1$~\cite{HP}.

For $q>q_+(r_+)>q_0(r_+)$ black holes unstable, but at the same time the
pure $AdS_4$ space-time is not dynamically preferred either.  The region
$q_-(r_+)< q<q_0(r_+)$ corresponds to the domain where $AdS_4$ is
entropically favored, but the black hole solution still corresponds to a
local minimum of the Gibbs action (\ref{gibbsa4}).

\begin{figure}
   \vskip0cm
   \centerline{\psfig{figure=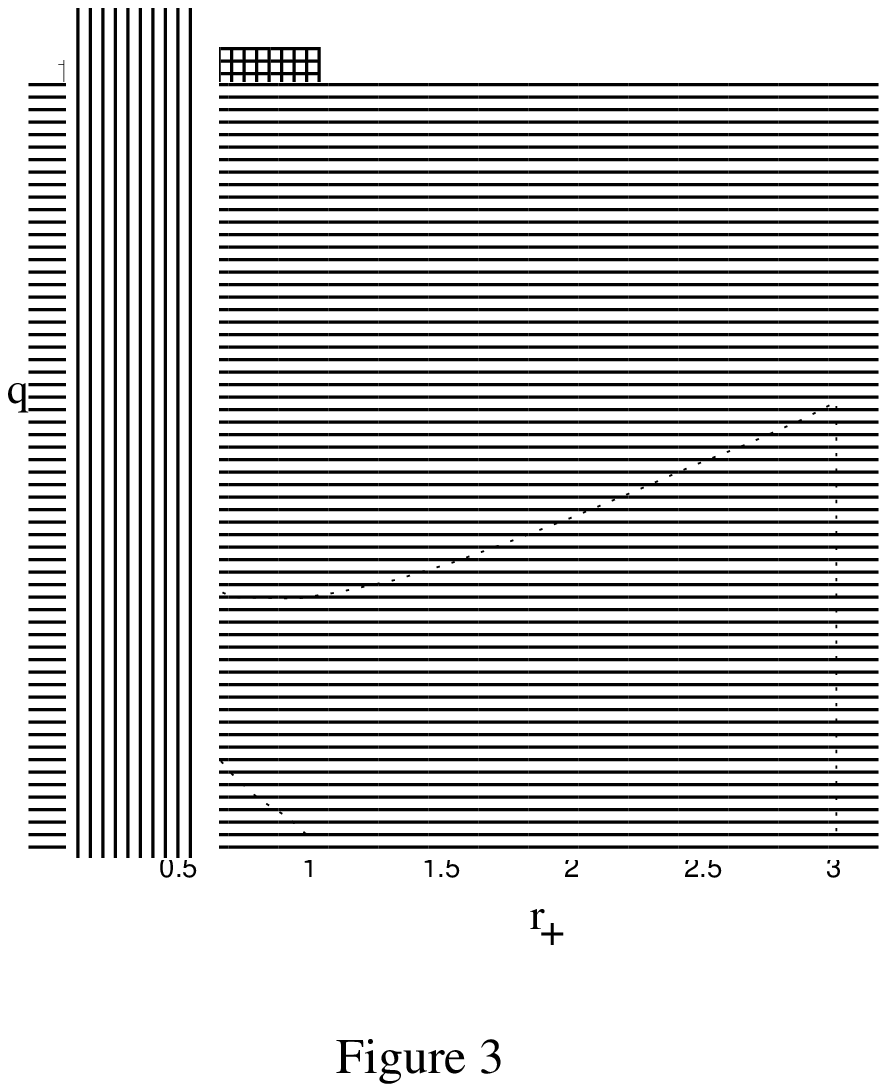,width=3in}}
   \vskip0cm
 \caption{The stability domains  as ${q}$ vs.~${{r_+}}$ 
  for the grand-canonical
 ensemble of  $D=4$
 R-charged black holes with one non-zero charge are given.  The vertically
shaded areas correspond to the regions where the $AdS_4$ is the preferred
solution and the horizontally shaded area correspond to the regions where
the black hole solutions are local minimum of the Gibbs action. The
checkered area is the domain of common overlap.  The
unshaded area corresponds to black holes that are unstable but nevertheless
favored over pure anti-de Sitter space.  Explicit powers of $L$ can be
restored by replacing $r_+$ with ${r_+\over L}$ and $q$ with ${{q}\over
L}$.
  }\label{figure3}
  \end{figure}

In Figure~3 we plot all these stability regions for  $q$ vs.~$r_+$. 
The vertically shaded area corresponds to the regions where the $AdS_4$ is the
preferred solution and the horizontally shaded area corresponds to the region
where the black hole solution is the local minimum of the Gibbs action.

\subsection{Canonical Ensemble and Stability Constraints}
Repeating the same analysis now for the  canonical ensemble yields the Helmholtz
action (\ref{helma}) which for 
 $D=4$ black holes  takes the following form:
 \begin{equation}
  I_H^{BH}={\beta\over{r_+}}\left[-r_+^4 +r_+^2(1+\sum_{i<j}^4q_iq_j)
  +r_+(\sum_{i=1}^4q_i+2\sum_{i<j<k}^4q_iq_jq_k)+3\prod_{i=1}^4q_i\right]\ , 
 \label{helmabh4}
  \end{equation}
 where $\beta$ is defined in (\ref{beta4}).
  $I_{H}^{BH}$ (\ref{helmabh4}) differs significantly  from  $I_{G}^{BG}$ 
   (\ref{gibbsabh4}).
   
 In the case of a single non-zero charge
 the local stability constraint leads to:
\begin{equation}
3r_+^5 + 6q r_+^4 + 2 r_+^3  +  7 q r_+^2  + 2 q^2 r_+ - r_+ - q\ge 0\ .
\label{canstab4}\end{equation}
The critical line now has only one branch:
\begin{equation}
q_{-}(r_+)={1\over{4r_+}}
\left(-6r_+^4-7r_+^2+1+\sqrt{36r_+^8+60r_+^6+21r_+^4-6r_+^2+1}\right) \ ,
\end{equation}
with  the region  $q\ge q_{-} (r_+)$   stable.

\begin{figure}
   \vskip0cm
   \centerline{\psfig{figure=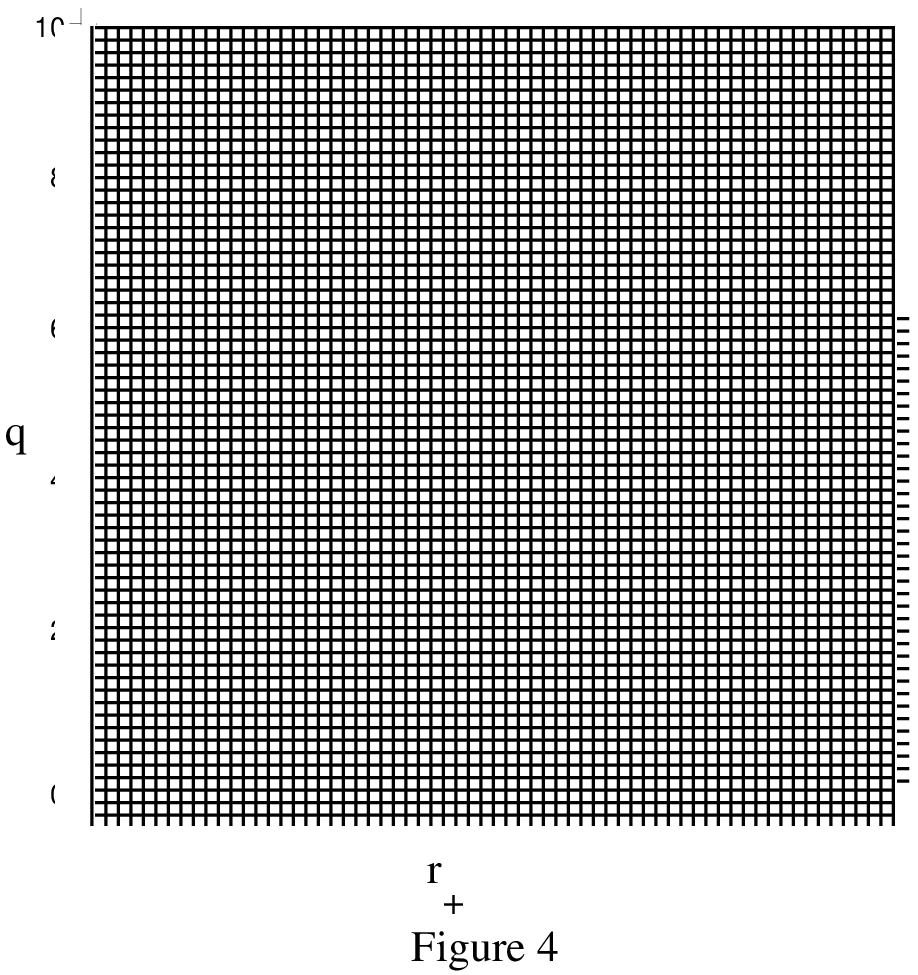,width=2.5in}}
   \vskip0cm
 \caption{The stability domains are plotted for ${q}$ vs. ${{r_+}}$ for the
canonical ensemble of $D=4$ R-charged black holes with one non-zero charge.
The vertically shaded areas indicates the regions where $AdS_4$ is the
preferred solution, and the horizontally shaded area shows the regions
where the black hole solutions are local minima of the Helmholtz action.
The checkered area is the domain of common overlap.  Explicit powers of $L$
can be restored by replacing $r_+$ with ${r_+\over L}$, and $q$ with
${{q}\over L}$.  }\label{figure4}
  \end{figure}

 The  phase transition between the black hole solution and the pure 
 $AdS_4$ solution  now takes place along the line
 \begin{equation}
 q_0(r_+)=r_+^3-r_+\ .  
 \end{equation}
 For $q>q_0(r_+)$ the black hole solution
   is the dynamically preferred solution.
 Pure 
$AdS_4$  is dynamically favored in the whole domain where the black hole is
thermodynamically unstable. 

In Figure 4 we plot the stability regions for $q$ vs. $r_+$.  The
vertically shaded area corresponds to the regions where $AdS_4$ is the
preferred solution, and the horizontally shaded area corresponds to the
region where the black hole solution is a local minimum of the Helmholtz
action.

\section{Phases of R-charged black holes in $D=7$}
\label{SevenDim}

We quote the black hole solution of $N=4$ gauged supergravity in $D=7$
(While the solution has not been derived from the actual $D=7$ Lagrangian
the structure of this solution is in close analogy\fixit{``implied'' seems
too strong a word here.  I do not see a logical implication: we are
proceeding by analogy} with the structure of the R-charged solution in
$D=4$ and $D=5$.):
 \begin{equation}
ds^2 = - {(H_1H_2)^{-{4/5}}} f dt^2 +  (H_1H_2)^{1/5}
 	\bigg(f^{-1} {dr^2} + 
	r^2 d\Omega_{5,k} \bigg) \ ,
	\end{equation}
where	  \\ 
\begin{equation}
f = k - {\mu \over r^4} + g^2 r^2 H_1H_2 \ , \ \ H_i=1+{{q_i}\over r^4}, \
(i=1, 2) \ . 
\label{f7}\end{equation}
We have chosen $G_7={\pi^2\over 4}$.  We are most interested in the case $k=1$. 
The  $q_i$   and ${\tilde q}_i$  are again related by (\ref{charges})
and the mass is  specified as:
\begin{equation}
M={\textstyle{5\over 4}} \mu +q_1 + q_2=
\textstyle{5\over 4}(r_+^4+r_+^6g^2H_1H_2)+q_1+q_2 \ , 
\end{equation}
where  again in the second equality the
radial coordinate at the outer horizon $r_+$  is defined as the largest
non-negative zero of $f(r)$ (\ref{f7}). 

The parameterization of the thermodynamic quantities in terms of 
of $q_i$ and $r_+$ yields the following expressions:
\begin{equation}
S= {A\over{4G_7}}=\pi r_+\sqrt{(r_+^4+q_1)(r_+^4+q_2)}\ ,
\label{ent7}\end{equation}
\begin{equation}
\beta={1\over T_H}={{\pi r_+^3\sqrt{
(r_+^4+q_1)(r_+^4+q_2)}}\over{3r_+^8+2r_+^6+r_+^4(q_1+q_2)-q_1q_2}} .
\label{beta7}\end{equation}
The physical charges are then determined in terms of $q_i$ and $r_+$
in the following way:
\begin{equation}
{\tilde q}_i^2=q_i(r_+^4+ q_i)\left[1 +{1\over r_+^2}
(r_+^4+q_j)\right] \ , \  (j\ne i)\ ,
\end{equation}
and the electric potentials at the outer horizon take the following  form:
\begin{equation}
\phi_i=A^i_t(r_+)={{\tilde q_i}\over {r_+^4 +q_i}} \ , \ (i=1,2)\ . 
\label{pot7}\end{equation}
 In the following we set $L \equiv {1\over g}= 1$. $L$, which is related to
the asymptotic cosmological constant $\Lambda=-{{15}\over L^2}$, can be
restored by replacing $r_+$ by ${r_+\over L}$ and $q_i$ by ${q_i\over
L^4}$.

\subsection{Grand-Canonical Ensemble  and 
Stability Constraints
 }

 $I_G$  for the 
 black hole solution, when expressed explicitly in
  terms of $q_i$ and $r_+$,
   takes the  form:
 \begin{equation}
  I_G^{BH}={\beta\over{2r_+^2}}\left[r_+^6-(r_+^4+q_1)(r_+^4+q_2)\right] \ ,
 \label{gibbsabh7}
  \end{equation}
where $\beta$ is determined by (\ref{beta7}). 

For one non-zero charge only,  the 
thermodynamic 
stability constraint takes the following form:
\begin{equation}
3r_+^8+r_+^6+2qr_+^4-2r_+^4+3qr_+^2-q^2\ge 0\ . 
\label{grandcanstab7}\end{equation}
The critical line  has two branches:
\begin{equation}
q_{\pm}(r_+)=r_+^2\left(r_+^2+\textstyle{3\over 2}+
\textstyle{1\over 2}\sqrt{16r_+^4+16r_+^2+1}\right)\ ,
\end{equation}
and the region of local stability is $q_{-}(r_+)<q<q_{+}(r_+)$.
 $q=0$   has the critical point at 
 $r_+=\sqrt{2\over 3}$.  Note also  that as $r_+\to 0$ the critical line 
 is pushed to $q_{\pm} \to 0$. 
 
 The  phase transition between the black hole solution and the pure 
 $AdS_7$  takes place at $q_0(r_+) = 0$, where 
 \begin{equation}
 q_0(r_+)=r_+^2-r_+^4 \ .  
 \end{equation}
 For $q>q_0(r_+)$ the black hole 
  is the dynamically preferred solution. As
 $r_+\to 0$ the critical line is pushed to $q_0\to 0$.  When $q_0=0$, the
Hawking-Page transition is at $r_+=1$.  In the region
$q>q_+(r_+)>q_0(r_+)$, the black hole solutions are unstable, but $AdS_7$
is not dynamically preferred either.

\begin{figure}
   \vskip0cm
   \centerline{\psfig{figure=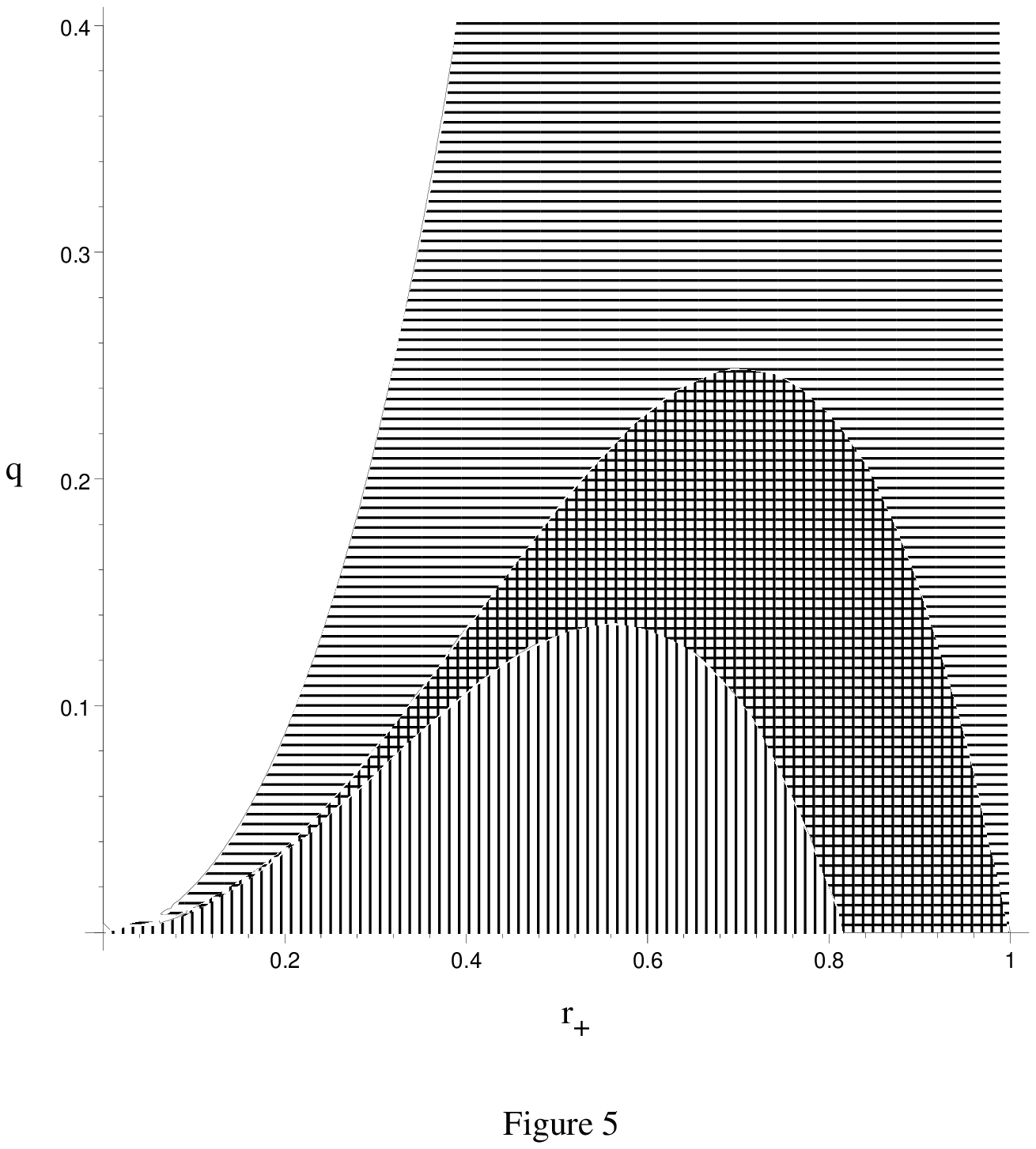,width=2.5in}}
   \vskip0cm
 \caption{The stability domains as ${q}$ vs.~${{r_+}}$
  for the grand-canonical
 ensemble of  $D=7$
 R-charged black holes with one non-zero charge.  The vertically shaded
areas correspond to the regions where the $AdS_7$ is the preferred solution
and the horizontally shaded area correspond to the regions where the black
hole solutions are local minimum of the Gibbs action. The checkered area is
the domain of common overlap.  The unshaded area is where neither black
holes nor anti-de Sitter space are stable.  Powers of $L$ can be restored
by replacing
 $r_+$  with ${r_+\over L}$  and $q$ with  ${{q}\over L^4}$.  }\label{figure5}
  \end{figure}
  
In Figure~5 the stability regions are shown in a plot of $q$ vs.~$r_+$.

\subsection{Canonical Ensemble and Stability Constraints}
Repeating the same analysis for the  canonical ensemble yields the Helmholtz
action (\ref{helma}) which for 
 the black hole solution  takes the following form:
 \begin{equation}
  I_H^{BH}={\beta\over{2r_+^2}}\left[-r_+^8 +r_+^6 +(3r_+^4+4r_+^2)
  (q_1+q_2) +7q_1q_2\right]\ , 
 \label{helmabh7}
  \end{equation}
 with  $\beta$ defined in (\ref{beta7}).
    
 In the case on single non-zero charge,
 the local stability constraint leads to:
\begin{equation}
3r_+^{10}+r_+^8-2r_+^6+12qr_+^6+17qr_+^4+4qr_+^2-3r_+^2q^2-2q^2\ge0\ .
\label{canstab7}\end{equation}
The critical line  has two branches:
\begin{equation}
q_{\pm}(r_+)= 
{{r_+^2}\over{2(2+3r_+^2)}}
\left(17r_+^2+4+12r_+^4+\sqrt{180r_+^8+444r_+^6+369r_+^4+120r_+^2+16}\right)\ , 
\end{equation}
with $q_-(r_+)\le q\le q_+(r_+)$ corresponding to
the stable region. Note that as $r_+\to 0$,  the critical lines are pushed to $q_{\pm} \to 0$.

\begin{figure}
   \vskip0cm
   \centerline{\psfig{figure=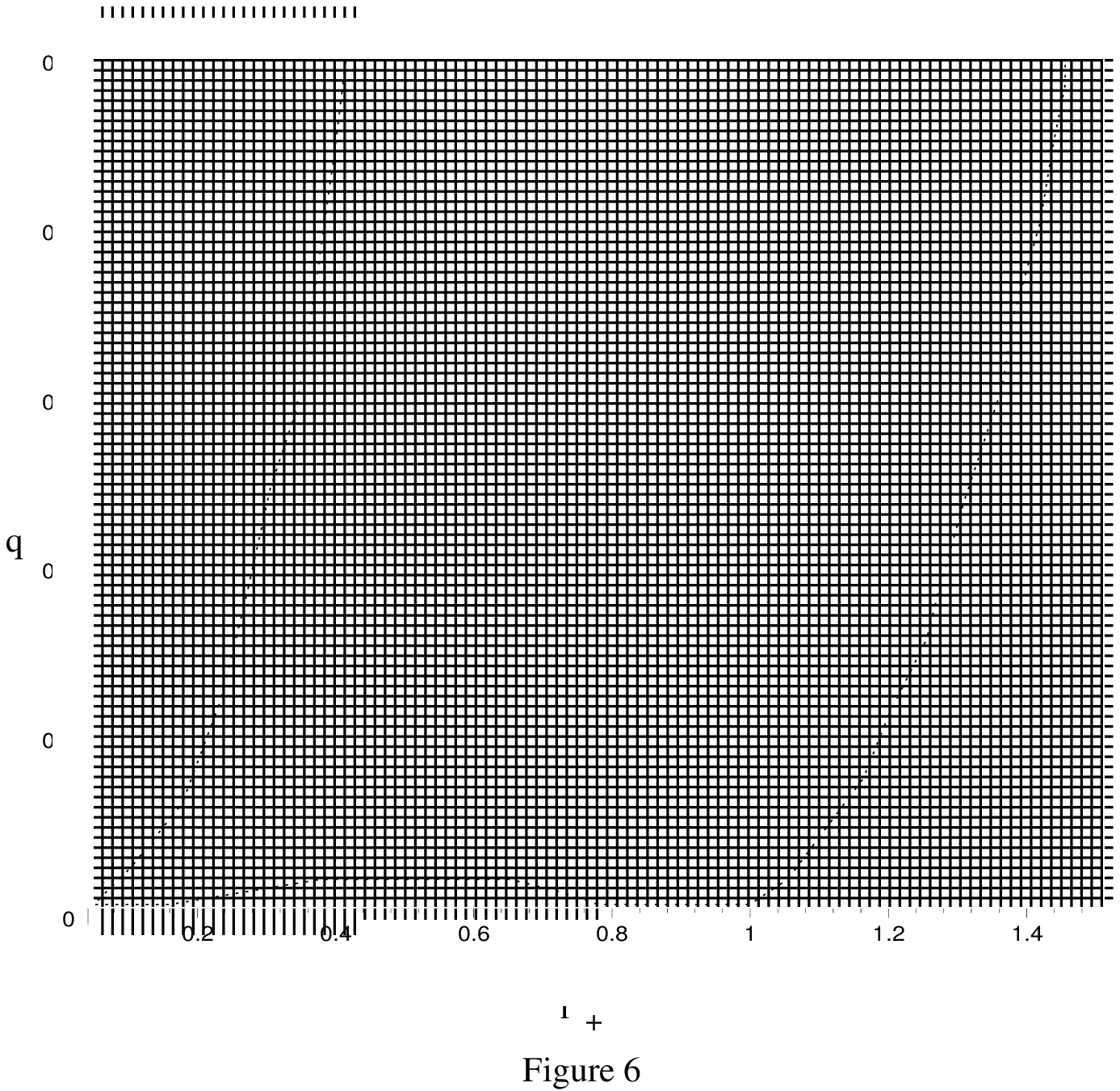,width=2.5in}}
   \vskip0cm
 \caption{The stability domains   ${q}$ vs. ${{r_+}}$ 
  for the canonical
 ensemble of  $D=7$
 R-charged black holes with one non-zero charge are given.  The vertically
shaded areas correspond to the regions where the $AdS_7$ is the preferred
solution and the horizontally shaded area correspond to the regions where
the black hole solutions are local minimum of the Helmholtz action.  The
checkered area is the domain of common overlap. The units of $L=1$ can be
restored by replacing
 $r_+$  with ${r_+\over L}$,  and $q$ with  ${{q}\over L^4}$.  }\label{figure6}
  \end{figure}

The Hawking-Page transition between the black hole solution and the pure
 $AdS_7$ solution is  now determined  by $q_0(r_+) = 0$, where 
 \begin{equation}
 q_0(r_+)={{r_+^6-r_+^4}\over {4r_+^2+3}}\ .  
 \end{equation}
 For $q>q_0(r_+)$ the black hole solution  is the dynamically preferred solution.
 As $r_+\to 0 $,  $q_0 \to 0$. 
In Figure~6  these stability domains are shown in a plot of $q$ vs.~$r_+$.

\section{ R-charged  black holes in  $D=5$ ($D=7$) and 
spinning M2 (M5)-branes} 
\label{Mbranes}

The Kaluza-Klein reduction described in Section~\ref{RCharge} of spinning
D3-branes to R-charged black holes in $AdS_5$ can be repeated for spinning
M2- and M5-branes~\cite{CvYou}, and it leads to R-charged black holes in $AdS_4$
and
$AdS_7$.  As in Subsection~\ref{equiv}, one can also obtain the precise
translation from the mass $M$, charges $\tilde{q}_i$, and the entropy of
large R-charged black holes in $AdS_4$ [$AdS_7$], to the energy $E$ above
extremality, the angular momenta $J_i$, and the entropy $S$ of 
near-extreme spinning M2-branes [M5-branes].

As a consequence of this equivalence, 
 the thermodynamic stability constraints 
derived for the general
spinning M2-branes [and M5-branes]  should
coincide with those of large R-charged black holes in $D=4$ [and $D=7$],
respectively.   For the case of only one non-zero charge
the    grand-canonical   ensemble 
(\ref{grandcanstab4}) [(\ref{grandcanstab7})] 
 and canonical ensemble (\ref{grandcanstab4}) [and (\ref{grandcanstab7})]
stability constraints reduce in the case  of large R-charged black 
holes ($r_+\gg 1$ and $q\gg 1$) to the
following   stability  constraints: 
\begin{eqnarray}
3r_+-2q\ge 0\ ,\ \  &D&=4,\  \hbox{grand-canonical}\ ,\\
r_++2q\ge 0\ , \ \   &D&=4,\  \hbox{canonical}\ ,
\label{large4}\end{eqnarray}

\begin{eqnarray}
[\ \ 3r_+^4-q\ge 0\ , \ \  &D&=7,\  \hbox{grand-canonical}\ ,\\
 r_+^8+4r_+^4q-q^2\ge 0 \ , \ \  &D&=7,\  \hbox{canonical}\ \ ]
\label{large7}\end{eqnarray}
 which agree precisely with the  respective stability 
constraints \cite{Cai,CvGu} for spinning  M2-branes [and M5-branes] with
 only one angular momentum turned on.

Finally, we would like to note a feature of the asymptotics of the
stability conditions for small and large $q$ which may be only
numerological coincidence, but seems to us intriguing.  There is an
exponent $\alpha_p$ which characterizes the scaling of the entropy $S$ of
near-extreme non-dilatonic $p$-branes with the number $N$ of branes: $S
\sim N^{\alpha_p}$.  For $p=3$ (the D3-brane), $\alpha_3 = 2$, which is
interpreted as evidence that there are indeed on the order of $N^2$ degrees
of freedom on the world-volume, as expected in a conformally invariant
gauge theory with gauge group $SU(N)$.  For $p=2$ (the M2-brane), $\alpha_2
= 3/2$.  For $p=5$ (the M5-brane), $\alpha_5 = 3$.  Cross-sections of
minimal scalars falling into any of these branes have the same
$N^{\alpha_p}$ scaling \cite{Itz,gPhD}.  We note for the record that
$\alpha_p = {d-3 \over d-5}$ where $d$ is one plus the number of spatial
dimensions perpendicular to the brane world-volume.
 
In the grand-canonical ensemble, the ratio of charge to mass where
stability is lost for large R-charged black holes in $D=4$ [$D=7$] is most
efficiently expressed as $\sqrt{r_+\over q}=\sqrt{2\over
3}=(\alpha_2)^{-1/2}$ [$\sqrt{r_+^4\over q}=
{1\over\sqrt{3}}=(\alpha_5)^{-1/2}$].  On the other hand, in the $q \to 0$
limit, the critical mass where stability is lost can be determined from
$r_+= {1\over\sqrt{3}}$ [$r_+=\sqrt{2\over 3}$] in $D=4$ [$D=7$].  Thus the
$D=4$ and $D=7$ results are in a peculiar numerological sense dual to one
another.  For the case of R-charged black holes in $D=5$, the large black
hole stability constraint and that of $q=0$ are self-dual:
$\sqrt{r_+^2\over q}={1\over\sqrt{2}}=(\alpha_3)^{-1/2}$ for large $q$, and
$r_+={1\over\sqrt{2}}$ for small $q$.

\section{Conclusions} 
\label{Conclude}

One of the motivations for this work was the question, ``What happens in a
quantum theory of gravity when a black hole becomes thermodynamically
unstable?''  String theory in principle provides us with a venue to address
this question.  In general, however, the black holes of which one has a
good string theory description are thermodynamically stable.  A possible
exception, first noted in \cite{Gubser}, is D3-branes with spin, by which
we mean angular momentum in a plane transverse to the world-volume.  These
objects, which we have shown to reduce to charged black holes in $AdS_5$,
can become unstable due to excessively large spin---or, in the
five-dimensional picture, excessive charge.

We had initially hoped to show that the charge-driven instability of black
holes in $AdS_5$ could be understood in terms of the Hawking-Page
transition: that is, for the black holes which were thermodynamically
unstable, there would be a transition to a gas of charged particles in pure
anti-de Sitter space.  To a first approximation, this transition occurs
when the Euclidean action of the black hole solution rises above the
Euclidean action for pure anti-de Sitter space subjected to the same
boundary conditions at spatial infinity.

This hope was only partly substantiated, and an intricate phase diagram
realized, through analytic treatments of the local thermodynamic stability
conditions and of the criterion for the Hawking-Page transition.  Holding
temperature and electric potential fixed, we found in particular that there
is a region of phase space where the black hole solution is locally
unstable (that is, it is only a saddle point and not a local minimum of the
Euclidean action), but where pure anti-de Sitter space has a still higher
action!  In this region, neither the black hole solution nor the gas of
particles in anti-de Sitter space can be the preferred state of the system.
The only option we can suggest at this point is that the preferred
configuration is not spherically symmetric with respect to rotations about
its center of mass in $AdS_5$, despite the fact that it has no angular
momentum in $AdS_5$---spin is angular momentum in the $S^5$ directions.
For instance, one might hope for a multi-black hole configuration.
Fragmentation of highly charged black holes would be a truly novel
phenomenon, but it does have some intuition behind it: near-extreme
D3-branes which spin too fast seem likely to split apart into chunks which
move out in the radial direction (which becomes a dimension in $AdS_5$ upon
Kaluza-Klein reduction).  The chunks could carry off some of the spin as
orbital angular momentum in the $S^5$ directions.  The ten-dimensional
geometry we are envisioning is neither stationary nor static, but it does
seem that its five-dimensional analog is a multi-black hole geometry.  We
hope that progress can be made in finding such a five-dimensional geometry,
or in proving that it cannot exist by some extension of the known
uniqueness theorems for black hole geometries.

Before we get carried away with the possibilities for quantum fission of
black holes, we should point out that the conclusions depend on whether
total charge or electric potential is held fixed.  If total charge is held
fixed, then in all the cases we considered ($D=4$ and $D=7$ as well as
$D=5$), for each black hole geometry which is locally thermodynamically
unstable, there is a pure anti-de Sitter geometry with the same total
charge and periodicity of Euclidean time which is entropically preferred in
the sense that it has smaller Euclidean action.  Thus for fixed total
charge, our initial hope is realized: black holes which get too highly
charged to exist in equilibrium with a heat bath undergo a Hawking-Page
transition.  As usual in first order phase transitions, it is also possible
for the black hole solution to represent a local minimum of the Euclidean
action (i.e.\ a local maximum of the entropy), to equilibrate with a heat
bath---in short to lead a stable existence---but nevertheless not to be the
global minimum of the action.  If one waits long enough, such
``meta-stable'' black holes should quantum tunnel to a gas of particles in
$AdS$ (plus perhaps a locally unstable black hole which then decays quickly
into the heat bath).

As we have indicated in Subsection~\ref{LocalGlobal}, the choice of
ensemble depends on physical motivation.  In the case of spinning branes,
where the world-volume is infinite and one is really considering a black
brane geometry where any part of the world-volume can act as a ``heat
bath'' for any other part, the grand canonical ensemble seems the
appropriate one for judging local stability.  For black holes in anti-de
Sitter space, the situation is less clear.  Our gravity intuition is that a
conserved charge should be held fixed when studying a localized object (the
black hole).  That is, we should use the canonical ensemble.  But if
gravity is wholly reflected in the gauge theory on the boundary, it seems
that, at least in $k\to 0^+$ limit where the world-volume of the gauge
theory is much larger than the cube of its inverse temperature, the same
arguments that applied to spinning branes tell us that we should be using
the grand canonical ensemble.  We have yet to resolve this issue to our
complete satisfaction.

Most of our results were obtained in the case of a single charge non-zero.
The number of independent charges is the rank of the gauged supergravity's
gauge group, or in the language of spinning branes the number of mutually
orthogonal planes perpendicular to the brane world-volume.  However, our
analysis lays the groundwork for an exploration of multiple non-zero
angular momenta.  In particular, we have obtained explicit closed forms for
the Euclidean actions, both with charges held fixed and with potentials
held fixed, for arbitrary numbers of charges up to the maximum number
allowed: three for the D3-brane, four for the M2-brane, and two for the
M5-brane.  Using these actions to locate the Hawking-Page transition is
straightforward: the way we have set up the calculations, the transition
takes place when the action is zero.  Evaluating the local stability
constraints may be feasible in the general case for the canonical ensemble,
but it is exceedingly calculationally burdensome in the grand-canonical
ensemble because determinants of matrices with up to five rows and columns
are involved.  In the case of large black holes we have been able to
evaluate these determinants.  This and other issues will be reported on in
\cite{CvGu}.

\vskip1cm
\noindent{\it Note added. \ \ } 
 As this paper was being completed we received the preprint \cite{Myers},
which overlaps with some of our results.

\acknowledgments
 We would like to thank J.~Distler, H.~L\"u, and A.~Strominger for
discussions.  The work of M.C.\ was supported in part by U.S. Department of
Energy Grant No.  DOE-EY-76-02-3071.  The work of S.S.G.\ was supported by
the Harvard Society of Fellows, and also in part by the National Science
Foundation under grant number PHY-98-02709 and by DOE grant
DE-FGO2-91ER40654.  Some of the results presented in this paper were
obtained with the help of Maple.

\newpage

\bibliography{rchargedbh}
\bibliographystyle{ssg}

\end{document}

%% file: lmacros.tex
\input umacros
\def\lbldef#1#2{\expandafter\gdef\csname #1\endcsname {#2}}
\def\eqn#1#2{\lbldef{#1}{(\ref{#1})}%
\begin{equation} #2 \label{#1} \end{equation}}
\def\eqalign#1{\vcenter{\openup1\jot
    \halign{\strut\span\TL & \span\TR\cr #1 \cr
   }}}
\def\eno#1{(\ref{#1})}

%% file: umacros.tex
\def\TL{\hfil$\displaystyle{##}$}
\def\TR{$\displaystyle{{}##}$\hfil}

\def\comment#1{}
\def\fixit#1{}

\def\mop#1{\mathop{\rm #1}\nolimits}

\def\vol{\mop{vol}}

\def\lsim{\mathrel{\mathstrut\smash{\ooalign{\raise2.5pt\hbox{$<$}\cr\lower2.5pt\hbox{$\sim$}}}}}
\def\gsim{\mathrel{\mathstrut\smash{\ooalign{\raise2.5pt\hbox{$>$}\cr\lower2.5pt\hbox{$\sim$}}}}}
\def\sqr#1#2{{\vcenter{\vbox{\hrule height.#2pt
         \hbox{\vrule width.#2pt height#1pt \kern#1pt
            \vrule width.#2pt}
         \hrule height.#2pt}}}}

\def\href#1#2{#2}  

%% file: rchargedbh.bbl
\begingroup\raggedright\begin{thebibliography}{10}

\bibitem{gkPeet}
S.~S. Gubser, I.~R. Klebanov, and A.~W. Peet, ``Entropy and temperature of
  black 3-branes,'' {\em Phys. Rev.} {\bf D54} (1996) 3915--3919,
  \href{http://xxx.lanl.gov/abs/hep-th/9602135}{{\tt hep-th/9602135}}.

\bibitem{JuanAdS}
J.~Maldacena, ``The Large N limit of superconformal field theories and
  supergravity,'' {\em Adv. Theor. Math. Phys.} {\bf 2} (1998) 231,
  \href{http://xxx.lanl.gov/abs/hep-th/9711200}{{\tt hep-th/9711200}}.

\bibitem{witHolTwo}
E.~Witten, ``Anti-de Sitter space, thermal phase transition, and confinement in
  gauge theories,'' {\em Adv. Theor. Math. Phys.} {\bf 2} (1998) 505,
  \href{http://xxx.lanl.gov/abs/hep-th/9803131}{{\tt hep-th/9803131}}.

\bibitem{WitSuss}
L.~Susskind and E.~Witten, ``The Holographic bound in anti-de Sitter space,''
  \href{http://xxx.lanl.gov/abs/hep-th/9805114}{{\tt hep-th/9805114}}.

\bibitem{Gubser}
S.~S. Gubser, ``Thermodynamics of spinning D3-branes,''
  \href{http://xxx.lanl.gov/abs/hep-th/9810225}{{\tt hep-th/9810225}}.

\bibitem{CvGu}
M. Cveti\v{c} and S. S. Gubser, UPR-826-T, forthcoming.

\bibitem{Cai}
R.-G. Cai and K.-S. Soh, ``Critical behavior in the rotating D-branes,''
  \href{http://xxx.lanl.gov/abs/hep-th/9812121}{{\tt hep-th/9812121}}.

\bibitem{HP}
S.~W. Hawking and D.~N. Page, ``Thermodynamics of black holes in anti-de Sitter
  space,'' {\em Commun. Math. Phys.} {\bf 87} (1983) 577.

\bibitem{romans}
L.~J. Romans, ``Supersymmetric, cold and lukewarm black holes in cosmological
  Einstein-Maxwell theory,'' {\em Nucl. Phys.} {\bf B383} (1992) 395--415,
  \href{http://xxx.lanl.gov/abs/hep-th/9203018}{{\tt hep-th/9203018}}.

\bibitem{BCS}
K.~Behrndt, M.~Cveti\v{c}, and W.~A. Sabra, ``Nonextreme black holes of
  five-dimensional N=2 AdS supergravity,''
  \href{http://xxx.lanl.gov/abs/hep-th/9810227}{{\tt hep-th/9810227}}.

\bibitem{Duff}
M.~J. Duff and J.~T. Liu, ``Anti-de Sitter black holes in gauged N = 8
  supergravity,'' \href{http://xxx.lanl.gov/abs/hep-th/9901149}{{\tt
  hep-th/9901149}}.

\bibitem{Yorketal}
H.~W. Braden, J.~D. Brown, B.~F. Whiting, and J.~James W.~York, ``Charged black
  hole in a grand canonical ensemble,'' {\em Phys. Rev.} {\bf D42} (1990)
  3376--3385.

\bibitem{Louko}
J.~Louko and S.~N. Winters-Hilt, ``Hamiltonian thermodynamics of the
  Reissner-Nordstrom anti- de Sitter black hole,'' {\em Phys. Rev.} {\bf D54}
  (1996) 2647--2663, \href{http://xxx.lanl.gov/abs/gr-qc/9602003}{{\tt
  gr-qc/9602003}}.

\bibitem{Peca}
C.~S. Peca and P.~S. Jose~Lemos, ``Thermodynamics of Reissner-Nordstrom anti-de
  Sitter black holes in the grand canonical ensemble,''
  \href{http://xxx.lanl.gov/abs/gr-qc/9805004}{{\tt gr-qc/9805004}}.

\bibitem{GubsKleTs}
S.~S. Gubser, I.~R. Klebanov, and A.~A. Tseytlin, ``Coupling constant
  dependence in the thermodynamics of N=4 supersymmetric Yang-Mills theory,''
  {\em Nucl. Phys.} {\bf B534} (1998) 202,
  \href{http://xxx.lanl.gov/abs/hep-th/9805156}{{\tt hep-th/9805156}}.

\bibitem{juanFive}
J.~M. Maldacena, ``Statistical entropy of near extremal five-branes,'' {\em
  Nucl. Phys.} {\bf B477} (1996) 168--174,
  \href{http://xxx.lanl.gov/abs/hep-th/9605016}{{\tt hep-th/9605016}}.

\bibitem{CvYou2}
M.~Cveti\v{c} and D.~Youm, ``Near BPS saturated rotating electrically charged
  black holes as string states,'' {\em Nucl. Phys.} {\bf B477} (1996) 449--464,
  \href{http://xxx.lanl.gov/abs/hep-th/9605051}{{\tt hep-th/9605051}}.

\bibitem{Larsen}
P.~Kraus, F.~Larsen, and S.~P. Trivedi, ``The Coulomb branch of gauge theory
  from rotating branes,'' \href{http://xxx.lanl.gov/abs/hep-th/9811120}{{\tt
  hep-th/9811120}}.

\bibitem{mahsch}
J.~Maharana and J.~H. Schwarz, ``Noncompact symmetries in string theory,'' {\em
  Nucl. Phys.} {\bf B390} (1993) 3--32,
  \href{http://xxx.lanl.gov/abs/hep-th/9207016}{{\tt hep-th/9207016}}.

\bibitem{witHolOne}
E.~Witten, ``Anti-de Sitter space and holography,'' {\em Adv. Theor. Math.
  Phys.} {\bf 2} (1998) 253, \href{http://xxx.lanl.gov/abs/hep-th/9802150}{{\tt
  hep-th/9802150}}.

\bibitem{hs}
M.~Henningson and K.~Skenderis, ``The Holographic Weyl anomaly,'' {\em JHEP}
  {\bf 07} (1998) 023, \href{http://xxx.lanl.gov/abs/hep-th/9806087}{{\tt
  hep-th/9806087}}.

\bibitem{BCSII}
K. Behrndt, M. Cveti\v{c}, and W. A. Sabra, unpublished.

\bibitem{CvYou}
M.~Cveti\v{c} and D.~Youm, ``Rotating intersecting M-branes,'' {\em Nucl.
  Phys.} {\bf B499} (1997) 253,
  \href{http://xxx.lanl.gov/abs/hep-th/9612229}{{\tt hep-th/9612229}}.

\bibitem{Itz}
N.~Itzhaki, J.~M. Maldacena, J.~Sonnenschein, and S.~Yankielowicz,
  ``Supergravity and the large N limit of theories with sixteen supercharges,''
  {\em Phys. Rev.} {\bf D58} (1998) 046004,
  \href{http://xxx.lanl.gov/abs/hep-th/9802042}{{\tt hep-th/9802042}}.

\bibitem{gPhD}
S. S. Gubser, {\it Dynamics of D-brane Black Holes}. Ph.D.~thesis, Princeton
  University, May, 1998.

\bibitem{Myers}
A.~Chamblin, R.~Emparan, C.~V. Johnson, and R.~C. Myers, ``Charged AdS black
  holes and catastrophic holography,''
  \href{http://xxx.lanl.gov/abs/hep-th/9902170}{{\tt hep-th/9902170}}.

\end{thebibliography}\endgroup
